\documentclass{aa}

\usepackage{booktabs} \usepackage{stfloats}

\usepackage{natbib} \usepackage[utf8]{inputenc}
\DeclareUnicodeCharacter{00A0}{~}
\DeclareUnicodeCharacter{00B1}{\ensuremath{\pm}}
\DeclareUnicodeCharacter{00B7}{\ensuremath{\cdot}}
\DeclareUnicodeCharacter{00D7}{\ensuremath{\times}}
\DeclareUnicodeCharacter{00F7}{\ensuremath{\div}}
\DeclareUnicodeCharacter{03B1}{\ensuremath{\alpha}} 
\DeclareUnicodeCharacter{03B2}{\ensuremath{\beta}}  
\DeclareUnicodeCharacter{20AC}{\euro} 
\DeclareUnicodeCharacter{2102}{\ensuremath{\mathbb{C}}} 
\DeclareUnicodeCharacter{2113}{\ensuremath{\ell}} 
\DeclareUnicodeCharacter{2115}{\ensuremath{\mathbb{N}}} 
\DeclareUnicodeCharacter{2119}{\ensuremath{\mathbb{P}}} 
\DeclareUnicodeCharacter{211A}{\ensuremath{\mathbb{Q}}} 
\DeclareUnicodeCharacter{211D}{\ensuremath{\mathbb{R}}} 
\DeclareUnicodeCharacter{2124}{\ensuremath{\mathbb{Z}}} 
\DeclareUnicodeCharacter{2190}{\ensuremath{\leftarrow}}
\DeclareUnicodeCharacter{2191}{\ensuremath{\uparrow}}
\DeclareUnicodeCharacter{2192}{\ensuremath{\rightarrow}}
\DeclareUnicodeCharacter{2193}{\ensuremath{\downarrow}}
\DeclareUnicodeCharacter{21A6}{\ensuremath{\mapsto}} 
\DeclareUnicodeCharacter{2200}{\ensuremath{\forall}}
\DeclareUnicodeCharacter{2202}{\ensuremath{\partial}}
\DeclareUnicodeCharacter{2205}{\ensuremath{\varnothing}}
\DeclareUnicodeCharacter{2207}{\ensuremath{\nabla}}
\DeclareUnicodeCharacter{2208}{\ensuremath{\in}}     
\DeclareUnicodeCharacter{2209}{\ensuremath{\not\in}} 
\DeclareUnicodeCharacter{2218}{\ensuremath{\circ}}   
\DeclareUnicodeCharacter{221D}{\ensuremath{\propto}}
\DeclareUnicodeCharacter{221E}{\ensuremath{\infty}}
\DeclareUnicodeCharacter{2220}{\ensuremath{\angle}}
\DeclareUnicodeCharacter{2260}{\ensuremath{\not=}}
\DeclareUnicodeCharacter{2264}{\ensuremath{\le}}
\DeclareUnicodeCharacter{2265}{\ensuremath{\ge}}
\DeclareUnicodeCharacter{2297}{\ensuremath{\otimes}}
\DeclareUnicodeCharacter{2299}{\ensuremath{\odot}}
\DeclareUnicodeCharacter{22C5}{\ensuremath{\cdot}}
\DeclareUnicodeCharacter{2715}{\ensuremath{\times}}
\DeclareUnicodeCharacter{2A2F}{\ensuremath{\times}}

\usepackage{amsmath,amsfonts,amssymb} 
\usepackage{mathtools} \usepackage[ruled,vlined]{algorithm2e}
\SetKwRepeat{Do}{do}{while}%
\usepackage{algpseudocode}

\SetKwComment{Comment}{$\triangleleft$ }{}

\newcommand{\AlgoStep}[2]{$\blacktriangleright$ \textbf{Step #1.} #2\;}

\usepackage{comment} 
\DeclareSymbolFont{bbold}{U}{bbold}{m}{n}
\DeclareSymbolFontAlphabet{\mathbbold}{bbold}

\usepackage{textcomp,mathrsfs}  
\usepackage{ifthen}

\usepackage[T1]{fontenc} \usepackage{lmodern} %
\usepackage{mathrsfs} \usepackage{stmaryrd} %
\usepackage{graphicx} 
\usepackage[dvipsname]{xcolor} \usepackage{tikz}

\definecolor{DarkBlue}{RGB}{0,69,134}      
\definecolor{Orange}{RGB}{255,66,14}       
\definecolor{Yellow}{RGB}{255,211,32}      
\definecolor{Green}{RGB}{87,157,28}        
\definecolor{DarkViolet}{RGB}{126,0,33}    
\definecolor{LightBlue}{RGB}{131,202,255}  
\definecolor{DarkGreen}{RGB}{49,64,4}      
\definecolor{LightGreen}{RGB}{174,207,0}   
\definecolor{Violet}{RGB}{75,31,111}       
\definecolor{Golden}{RGB}{255,149,14}      
\definecolor{Red}{RGB}{197,0,11}           
\definecolor{Blue}{RGB}{0,132,209}         

\usepackage[colorlinks=true,allcolors=DarkBlue]{hyperref}

\usepackage[normalem]{ulem}

\usepackage{xspace}

\usepackage[squaren,Gray]{SIunits}

\usepackage{numprint}

\usepackage{multirow} \usepackage{threeparttable}

\DeclarePairedDelimiterX{\Paren}[1]{(}{)}{#1}
\DeclarePairedDelimiterX{\Brace}[1]{\{}{\}}{#1}
\DeclarePairedDelimiterX{\Brack}[1]{[}{]}{#1}
\DeclarePairedDelimiterX{\Abs}[1]{\rvert}{\lvert}{#1}
\DeclarePairedDelimiterX{\Norm}[1]{\lVert}{\rVert}{#1}
\DeclarePairedDelimiterX{\Avg}[1]{\langle}{\rangle}{#1}
\DeclarePairedDelimiterX{\Round}[1]{\lfloor}{\rceil}{#1}
\DeclarePairedDelimiterX{\Floor}[1]{\lfloor}{\rfloor}{#1}
\DeclarePairedDelimiterX{\Ceil}[1]{\lceil}{\rceil}{#1}
\DeclarePairedDelimiterX{\Inner}[2]{\langle}{\rangle}{#1,#2}

\DeclareMathOperator*{\argmin}{arg\,min}

\DeclareMathOperator{\Diag}{diag}

\DeclareMathOperator{\Trace}{tr}

\DeclarePairedDelimiterXPP{\Expect}[1]{\mathbb{E}}(){}{#1}
 
\newcommand*{\estim}[1]{\widehat{#1}}

\newcommand*{\Set}[1]{\mathbb{#1}} \newcommand*{\Reals}{\Set{R}}
 
\newcommand*{\V}[1]{\boldsymbol{#1}}   
\newcommand*{\M}[1]{\mathbf{#1}}       
\newcommand*{\TransposeLetter}{\hspace*{-.25ex}\top\hspace*{-.25ex}}
\newcommand*{\T}{^{\TransposeLetter}} 

\DeclareFontFamily{U}{mathx}{\hyphenchar\font45}
\DeclareFontShape{U}{mathx}{m}{n}{<-> mathx10}{}
\DeclareSymbolFont{mathx}{U}{mathx}{m}{n}
\DeclareMathAccent{\widebar}{0}{mathx}{"73}

\newcommand{\PACO}{{\texttt{PACO}}\xspace}  
\newcommand{\REXPACO}{{\texttt{REXPACO}}\xspace}

\newcommand{\hide}[1]{} \newcommand{\DisjointPatches}{\ensuremath{\mathbb{P}}}

\newcommand*{\from}{\,{:}~} 

\newcommand{\SampleCov}{\widetilde{\M C}} \newcommand{\ShrunkCov}{\widehat{\M
		C}}

\author{Olivier Flasseur\inst{\ref{inst1}} \and Samuel Th\'{e}\inst{\ref{inst1}}
	\and Lo\"{i}c Denis\inst{\ref{inst2}} \and \'{E}ric
	Thi\'{e}baut\inst{\ref{inst1}} \and Maud Langlois\inst{\ref{inst1}}}

\institute{ Université de Lyon, Université Lyon1, ENS de Lyon, CNRS, Centre de
	Recherche Astrophysique de Lyon UMR 5574, F-69230, Saint-Genis-Laval, France
	\label{inst1} \\ \email{surname.name@univ-lyon1.fr} \and Université de Lyon,
	UJM-Saint-Etienne, CNRS, Institut d Optique Graduate School, Laboratoire Hubert
	Curien UMR 5516, F-42023, Saint-Etienne, France \label{inst2} \\
	\email{surname.name@univ-st-etienne.fr} }

\date{\today}

\title{REXPACO: An algorithm for high contrast reconstruction \\of the circumstellar environment by angular differential imaging}

\titlerunning{REXPACO -- Reconstruction of EXtended features by PAtch
	COvariances}

\abstract{%
	Direct imaging is a method of choice for probing the close
	environment of young stars. Even with the coupling of adaptive optics
	and coronagraphy, the direct detection of off-axis sources such as circumstellar
	disks and exoplanets remains challenging due to the required high contrast and
	small angular resolution. Angular differential imaging (ADI) is an
	observational technique that introduces an angular diversity to help
	disentangle the signal of off-axis sources from the residual signal of the
	star in a post-processing step.%
}{%
	While various detection algorithms have been proposed in the last decade to
	process ADI sequences and reach high contrast for
	the detection of point-like sources, very few methods are available to
	reconstruct meaningful images of extended features such as
	circumstellar disks. The purpose of this paper is to describe a new
	post-processing algorithm dedicated to the reconstruction of the
	spatial distribution of light (total intensity) received from
	off-axis sources, in particular from circumstellar disks.%
}{%
	Built on the recent \PACO algorithm dedicated to the detection of point-like
	sources, the proposed method is based on the local learning of patch
	covariances capturing the spatial fluctuations of the stellar
		leakages. From this statistical modeling, we develop a regularized image
	reconstruction algorithm (\REXPACO) following an inverse problems
	approach based on a forward image formation model of the off-axis
	sources in the ADI sequences.%
}{%
	Injections of fake circumstellar disks in ADI sequences from the
	VLT/SPHERE-IRDIS instrument show that both the morphology and the photometry of
	the disks are better preserved by \REXPACO compared to standard post-processing
	methods such as cADI. In particular, the modeling of the spatial covariances
	proves useful in reducing typical ADI artifacts and in better disentangling
	the signal of these sources from the residual stellar contamination. The
	application to stars hosting
	circumstellar disks with various morphologies confirms the ability of \REXPACO
	to produce images of the light distribution with reduced artifacts. Finally, we
	show how \REXPACO can be combined with \PACO to disentangle the signal of
	circumstellar disks from the signal of candidate point-like sources.%
}{%
	\REXPACO is a novel post-processing algorithm for reconstructing
	images of the circumstellar environment from high contrast ADI
	sequences. It produces numerically deblurred
	images and exploits the spatial covariances of the stellar
	leakages and of the noise to efficiently
	eliminate this nuisance term. The processing is fully unsupervised, all tuning
	parameters being directly estimated from the data themselves. %
}

\keywords{Techniques: image processing - Techniques: high angular resolution -
	Methods: statistical - Methods: data analysis.}

\begin{document}

	\maketitle

	\section{Introduction} \label{sec:introduction}

	\begin{figure*}[htbp] \centering
			\includegraphics[width=\textwidth]{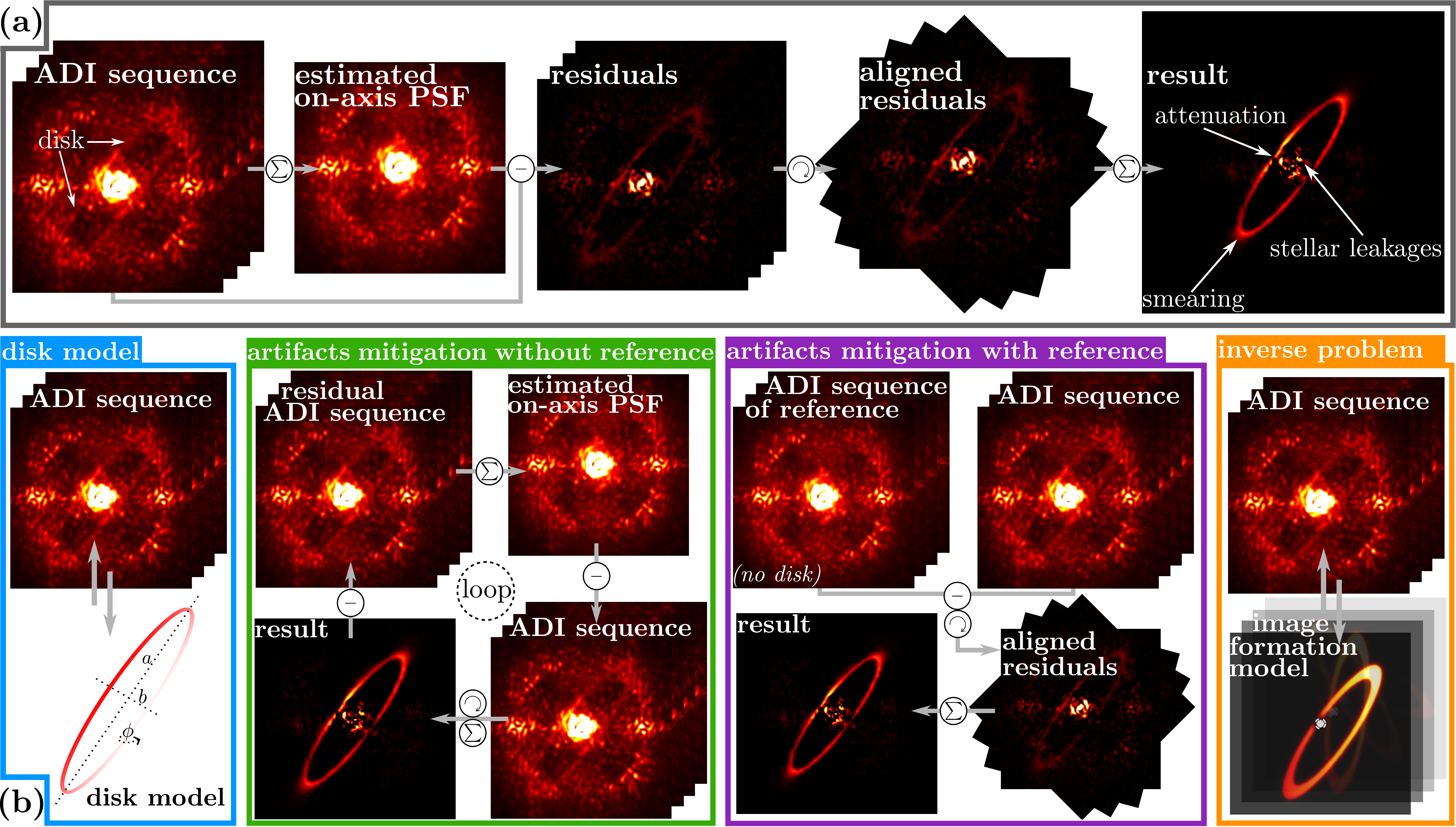} \caption{%
			Schematic representation of the main steps followed
			by the state-of-the-art post-processing methods for ADI
			sequences with circumstellar disks. (a) Principle of the classical approach
			based on the estimation and subtraction of an on-axis PSF. (b) Four main
			categories of algorithms designed to reduce reconstruction artifacts seen in
			(a): using a physics-based model of the disk, mitigating artifacts directly
			from the ADI sequence of interest or using an additional sequence of
			reference, or formalizing the reconstruction task as an inverse
			problem. \REXPACO falls into the last category: methods based on an inverse problems approach.}
			\label{fig:algos_fullfig}
	\end{figure*}

	Circumstellar disks are at the heart of the planet formation
	processes. Direct imaging in the near infrared is a unique method for addressing
	this question because planets can be detected together with the disk
	environment. Despite the very promising results from direct imaging, a small
	fraction of the expected circumstellar disks have been unveiled and
	characterized using high contrast direct imaging in the near infrared
	\citep{Esposito2020, Garufi2020, Langlois2020}. Recent protoplanetary and
	debris disk studies performed in intensity or in polarimetry have focused on
	morphological disk characteristics such as the presence of spirals
	\citep{Benisty2015, Ren2018, muro2020shadowing}, asymmetries, warps
	\citep{Dawson2011, Kluska2020} and gaps \citep{Boekel2017} that could be
	signposts for the presence of exoplanets. In this context, protoplanetary
	disks
	allow unique studies of the exoplanet-disk interactions
	\citep{keppler2018discovery, haffert2019two, mesa2019vlt}. In order to
	understand the physical processes governing these disks, it is
	essential to reconstruct their surface-brightness or scattering phase
	function but also to
	disentangle, in a post-processing
    step, the disks and the exoplanets.

	\medskip

	\noindent From an observational point of view, the direct detection of
	circumstellar disks requires: (i) high angular resolution to resolve the close
	environment of stars (the typical size of disks is smaller than 1 arcsecond) and 
	(ii) high contrast to detect the faint light scattered by disks (in the infrared, an effective contrast better than $10^5$ relative to the host star is typically required).
	Among the available ground-based observing facilities, the Spectro-Polarimetry High-contrast Exoplanet Research
	instrument \citep[SPHERE;][]{beuzit2019sphere} operating at the Very Large Telescope
	(VLT) offers such capabilities. It produces angular
	differential imaging \citep[ADI;][]{marois2006angular} sequences using the
	pupil-tracking mode of the telescope as a means to enhance the achievable
	contrast.
	The presence of strong residual stellar leakages from the coronagraph, in the form of so-called
	speckles, is the
	current main limitation of the reachable contrast. A post-processing step
	is required to combine the images of the ADI sequence and
	cancel out most of the nuisance
	component\footnote{We refer to the term ``nuisance component'' to designate the
		contribution of light sources other than the object of interest, as
		well as different sources of noise.} while
	preserving the signal from the off-axis
	sources\footnote{Since the signals from circumstellar disks
	and
			point-like sources come from directions other than the optical
			axis, we use
			the term ``off-axis sources'' to refer indifferently to these two types
			of objects.}.

	\medskip

	\noindent There is a large variety of methods dedicated to the post-processing
	of ADI sequences (see for example \cite{pueyo2018direct} for a review). Even though
	these methods are primarily designed for the detection of unresolved point-like
	sources, they are also widely used to process observations of disks
	\citep{milli2012impact} and share a common framework with most of the existing
	post-processing methods designed for disks. The common
	principle is to estimate a reference
	image\footnote{\samepage Since the nuisance component is mainly due to
		light coming from the optical axis, we refer to the term ``on-axis
			point-spread function'' (in short, ``on-axis PSF'') to designate the
		estimated reference image of the nuisance component.} of the nuisance
	component (see Fig. \ref{fig:algos_fullfig}(a) for a schematic
	illustration). This reference image is subtracted to each image of the
	sequence. The resulting residual images are aligned along a common
	direction to spatially
	superimpose the signals from the off-axis sources, and are added to each others.
	More sophisticated methods follow an inverse problems framework. In particular, the recent \PACO
	algorithm \citep{flasseur2018exoplanet, flasseur2018unsupervised,
		flasseur2018SPIE} jointly learns the average nuisance component and its
	fluctuations by estimating spatial covariances. Since the nuisance
	component is highly nonstationary, the underlying statistical model is
	estimated at a local scale of small patches.

	Most of these standard ADI post-processing algorithms are not suited to detect
	and reconstruct the light distribution of circumstellar disks. Their major
	drawback is the presence, in the output images, of strong artifacts that take the
	form of partial replicas, the removal of extended smooth components, smearing, and
	nonuniform attenuations due to the so-called self-subtraction phenomenon. As a direct consequence, both the
	morphology and the photometry of the disks are strongly corrupted.
	Very few specific solutions have been developed to tackle these
	limitations in the ADI processing of circumstellar disks. The existing approaches
	can be classified into four categories (see Fig.~\ref{fig:algos_fullfig}(b) for
	a schematic diagram of their principles).

	\begin{figure*}[htbp] \centering
		\includegraphics[width=\textwidth]{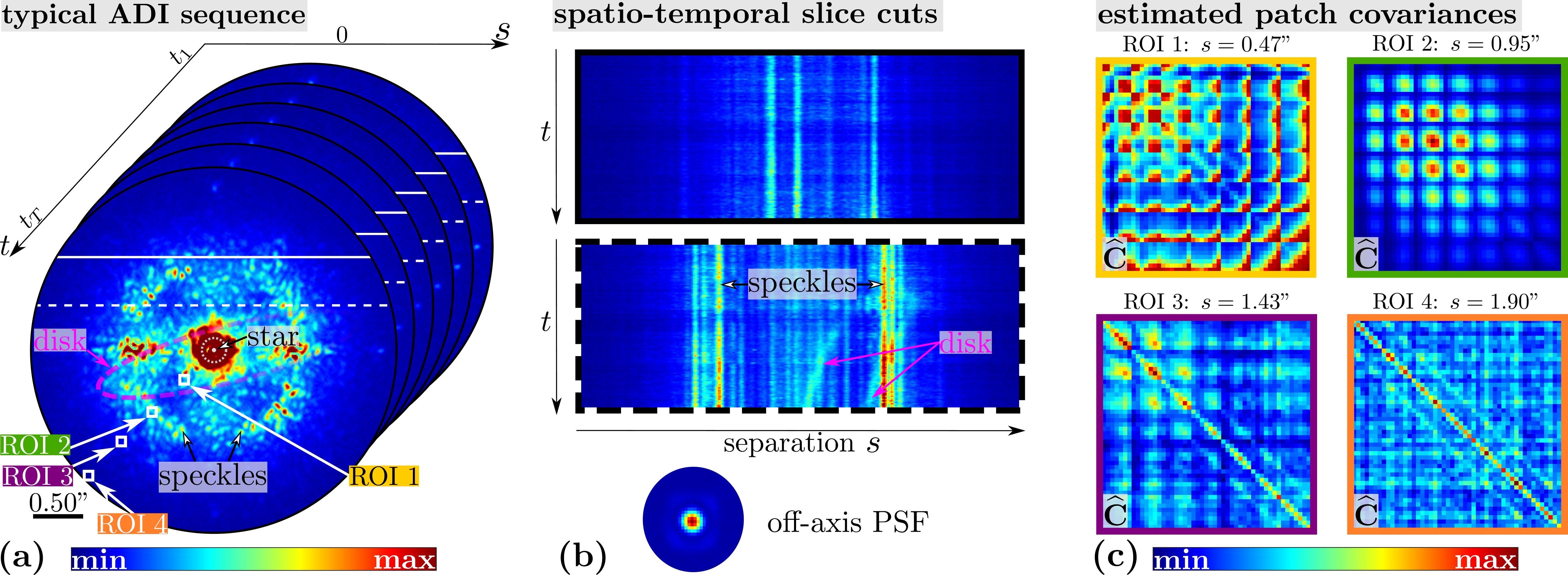}
		\caption{Typical ADI
			sequence from the VLT/SPHERE-IRDIS instrument: (a) Examples of
			temporal frames; (b) two spatio-temporal slices cut along the solid and dashed
			lines of (a) emphasizing the spatial variations of the structure of the
			signal; (c) estimated covariance matrices for four regions of interest (ROIs)
			at different angular separations $s$ with the host star. Data set: HR
			4796A, see Sect.~\ref{sec:datasets_description} for observing conditions.}
		\label{fig:disk_data_fullfig}
	\end{figure*}
	
	A first family of methods introduce a forward-backward loop embedding a
	physics-based model of the disk \citep{milli2012impact, esposito2013modeling, mazoyer2020diskfm}. In a nutshell, the parameters
	constraining the disk morphology and flux are optimized by minimizing the
	residuals produced by the selected post-processing algorithm from the ADI
	sequence free from the current estimation of the disk contribution. This method
	is however limited as it requires a physics-based model of the disk
	that is sufficiently simple to be numerically optimized and yet sufficiently flexible to accurately model the actual
	disk.

	A second category of methods attempt to avoid self-subtraction by reducing the
	disk contribution in the estimation of the reference on-axis PSF.
	For instance, \citet{pairet2018reference} have proposed a modified version of the PCA-based
	post-processing algorithm that iteratively removes the estimated disk component
	to improve the estimation of the on-axis PSF. To limit self-subtraction,
	\citet{ren2020using} have proposed a data-imputation strategy: The areas of the
	field of view impacted by the disk are considered as missing-data based on
	prior knowledge of the disk location and replaced by regions free from any disk
	contribution.

	So-called reference differential imaging \citep[RDI;][]{Gerard2016, wahhaj2021search} is a
	third class of approaches that require an additional ADI sequence of a reference
	star that hosts no known off-axis sources
	 to estimate the stellar contamination. The quality of the
	post-processing highly depends both on the selection of the reference star,
	and on the similarity of the observing conditions \citep{ruane2019reference}.
	\cite{ren2018non} also use a reference ADI sequence to build a
	nonnegative decomposition (via nonnegative matrix factorization) of
	the nuisance component.

	Finally, a fourth class of methods is emerging and states the reconstruction of
	the object of interest as an inverse problem. It consists in jointly estimating
	the nuisance term, the disk and, possibly, point-like sources given the data
	and low-complexity priors adapted to each kind of contribution
	\citep{pairet2019iterative, pairet2020mayonnaise}. The method we propose in this
	paper belongs to this family.

	\medskip

	\noindent Based on the previous analysis of the astrophysical needs and of the
	state-of-the-art post-processing methods for circumstellar disk detection and
	reconstruction, the desirable properties of post-processing algorithms may be
	listed as follows: (i) detection of disks at high contrasts, (ii) reconstruction
	of a physically plausible image of the flux distribution with limited
	processing artifacts, and (iii) ability to disentangle the spatially extended
	contribution of disks from the spatially localized contribution of point-like
	sources.
	
	In this paper, we attempt to address these three points by proposing a new
	post-processing method to extract the light distribution of extended features
	like circumstellar disks from ADI sequences. To get rid of the nuisance term, we formalize the
	reconstruction as an inverse problem where all the contributions appear explicitly and where the measured instrumental PSF is
	taken into account. As a result, the estimation of the object of interest is nearly free from contamination by the stellar leakages and
	from blurring by the off-axis PSF. The method proposed in this paper, named
	\REXPACO (for \textit{Reconstruction of EXtended features by PAtch
		COvariances}), introduces dedicated constraints for each component
	to help disentangle the object of interest from the nuisance term. For the object of interest, these
	constraints take the form of specific regularizations favoring the smoothness of
	extended structures (e.g., disk-like component) or the sparsity of point-like
	sources. For the nuisance term, we consider the same statistical model as with
	\PACO's algorithm \citep{flasseur2018exoplanet, flasseur2018unsupervised,
		flasseur2018SPIE}, initially designed for exoplanet detection, but which has
	also proven very effective for the recovery of extended patterns in holographic
	microscopy \citep{flasseur2019expaco}. The power of \PACO statistical model
	resides in its ability to capture the local behavior of the stellar leakages
	and the noise through the spatial covariances. Another advantage is
	that this model is directly learned from the ADI sequence and requires no other
	specific calibration data. Similarly, \REXPACO is an unsupervised
	algorithm: All tuning parameters are automatically estimated from the data. The
	framework implemented by \REXPACO is rather flexible. As an example, \REXPACO
	can be coupled with \PACO to better estimate off-axis point-like sources.

	The paper is organized as follows. In Sect.~\ref{sec:principle}, we formalize
	\REXPACO algorithm. In Sect.~\ref{sec:results}, we
	illustrate the results obtained by \REXPACO, both with semisynthetic data and with real
	observations of known circumstellar disks from the InfraRed Dual
	Imaging Spectrograph \citep[IRDIS;][]{dohlen2008infra, dohlen2008prototyping}
	of the VLT/SPHERE instrument. Section \ref{sec:results_joint} presents results
	from the combination of \REXPACO with the \PACO algorithm to unmix the
	spatially extended contribution of disk from the point-like contribution of
	exoplanets. Section~\ref{sec:conclusion} summarizes our contributions.

	\section{\REXPACO: Reconstruction of EXtended features by PAtch COvariances
		modeling} \label{sec:principle}

	This section describes \REXPACO, the method we propose to recover the spatial
	distribution $\V x$ of the light flux of off-axis sources from an ADI sequence
	$\V{r}\in\mathbb{R}^{N\,T}$ formed by $T$ images of $N$ pixels each.
	Table~\ref{tab:notation_reminder} summarizes the main notations used throughout
	the paper.%

	\subsection{Inverse problems formulation} \label{sec:general_framework}

		\begin{table}[t] \centering \caption{Summary of the main notations.}
			\begin{tabular}{@{}c@{}c@{\hspace*{1ex}}l@{}} \toprule \textbf{Not.}\; &
				\textbf{Range}\; &\textbf{Definition} \\ \midrule
				\multicolumn{3}{c}{$\triangleright$ Constants and related indexes} \\
				\midrule
				$K$ & $\mathbb{N}$ & number of pixels in a patch \\
				$M$ & $\mathbb{N}$ & number of pixels in a reconstructed image \\
				$N$ & $\mathbb{N}$ & number of pixels in a temporal frame \\
				$T$ & $\mathbb{N}$ & number of temporal frames \\
				$n$ & $\llbracket 1,N \rrbracket$ & pixel index \\
				$t$ & $\llbracket 1,T \rrbracket$ & temporal index \\ \midrule
				\multicolumn{3}{c}{$\triangleright$ Data quantities} \\ \midrule
				$\V r$ & $\mathbb{R}^{NT}$ & ADI sequence (observed data) \\
				$\V f$ & $\mathbb{R}^{NT}$ & nuisance component \\
				$\V x$ & $\mathbb{R}_+^{M}$ & light distribution of off-axis sources \\
				\midrule
				\multicolumn{3}{c}{$\triangleright$ Operators} \\ \midrule
				$\M A$ & $\mathbb{R}^{NT\times M}$ & off-axis PSF model: $\M A=\M V \, \M
				H \, \M \Gamma \,\M Q$ \\
				$\M Q$ & $\mathbb{R}^{MT\times M}$ & field rotation \\
				$\M \Gamma$ & $\mathbb{R}^{MT\times MT}$ & coronagraph attenuation \\
				$\M H$ & $\mathbb{R}^{MT\times MT}$ & convolution by off-axis PSF \\
				$\M V$ & $\mathbb{R}^{NT\times MT}$ & field of view cropping \\
				$\M P$ & $\mathbb{R}^{K\times N}$ & patch extractor \\ \midrule
				\multicolumn{3}{c}{$\triangleright$ Estimated quantities} \\ \midrule
				$\widehat{\V m}$ & $\mathbb{R}^{N}$ & temporal mean of $\V f$ \\
				$\SampleCov$ & $\mathbb{R}^{K\times K}$ & empirical spatial covariance of
				$\V f$ \\
				$\widehat{\M C}$ & $\mathbb{R}^{K\times K}$ & shrunk spatial covariance
				of $\V f$ \\
				$\widehat{\M \Omega}$
				& $\mathbb{R}^{NT}$ & statistics of the nuisance component $\V f$ \\
				$\widehat{\V x}$ & $\mathbb{R}_+^{M}$ & reconstructed distribution of
				light\\ \bottomrule
			\end{tabular}
			\label{tab:notation_reminder}
		\end{table}

	\REXPACO follows an inverse problems formulation: It computes an estimate
	$\estim{\V x}$ from the sequence of observations $\V r$ by fitting a model of
	the data under given constraints. Since all observations $\V{r}$ are
	collected during a single night, the distribution of interest $\V x$ can be
	considered time-invariant and described by a vector $\V{x}\in
	\mathbb{R}_{+}^M$, where $M$ is the number of pixels of the
	reconstructed field of view and where $\mathbb{R}_{+}$ is the set of
	nonnegative real numbers to account that the intensity is necessarily
	nonnegative everywhere. The ADI data sequence $\V r$ results from the
	contribution of two components as expressed by:
	\begin{equation}
		\V r = \M A \, \V x + \V f\,,
		\label{eq:additive_model}
	\end{equation}
	where $\M A$
	is the linear operator ($\mathbb{R}^M \to \mathbb{R}^{N T}$) modeling the
	effects of the off-axis instrumental point spread function (off-axis PSF), that is to say the non-coronagraphic image of the star, and
	$\V f\in\mathbb{R}^{NT}$ is a nuisance component accounting for
	the residual stellar light in the presence of aberrations (that is, the speckles)
	 {and for the contribution of noise. Due to the temporal
			variation of aberrations and to the noise, the nuisance component $\V f$ is
			the
			result of some random process. The quality of the statistical modeling of this
			process is crucial for the inversion of Eq.~(\ref{eq:additive_model}), and it
		represents a key element of \REXPACO. Figure~\ref{fig:disk_data_fullfig}(a)
		displays an example of an ADI sequence from the star HR~4796A observed by
		VLT/SPHERE-IRDIS. HR~4796A is surrounded by a bright inclined debris disk (see
		Sect.~\ref{sec:datasets_description} for a more complete description of its
		astrophysical properties). We selected this data set to illustrate the rationale
		of the different ingredients of \REXPACO. Part~(a) of
		Fig.~\ref{fig:disk_data_fullfig} shows that the bright disk of HR~4796A can
		only be marginally detected by a close visual inspection of a given frame of
		the sequence. This is due to the very large contrast between the star
		and its environment: The contribution of the component of interest $\M A\, \V
		x$ is typically fainter than the nuisance component $\V f$ by one to four
		orders of magnitude. This high contrast sets the gain that must be achieved by
		the post-processing method in order to recover the component of interest $\V x$
		from the data. Part~(b) of Fig.~\ref{fig:disk_data_fullfig} shows two
		spatio-temporal slices extracted at two different locations (along the solid
		line: far from the host star, along the dashed line: closer to the host star)
		and illustrates the nonstationarity of the nuisance component. Near the star
		(i.e., at angular separations smaller than about 1.5 arcsecond), the nuisance
		component is dominated by speckles. At larger angular separations, the nuisance
		component mainly results from a combination of photon noise, thermal background flux and
		detector readout. Part~(c) of Fig.~\ref{fig:disk_data_fullfig} shows the
		empirical spatial covariances of the data computed by temporal averaging
		through the ADI sequence in different regions of interest (ROIs). This figure
		not only reveals that the covariances of the fluctuations are spatially
		nonstationary but also that the covariances are non-negligible in a small
		neighborhood. In a nutshell, the nuisance component $\V f$ has
		strong spatial correlations and its distribution is spatially nonstationary and
		fluctuates over time, making the extraction of the
		object $\V x$ all the more difficult. 
		We introduce in Sect.~\ref{sec:learning_statistics} a model
		of the probability density function $\text{p}(\V f)$ of the nuisance component
		$\V f$ that is suitable to describe the statistical behavior of these
		fluctuations.

		We formalize the reconstruction of the object $\V x$ from the data $\V r$ as
		an inverse problem: We seek the light distribution $\widehat{\V x}$
		that would produce observations, given the instrumental model $\M A$,
		statistically close to the actual measurements. Image reconstruction methods
		based on the resolution of inverse problems are generally expressed in the
		form of an optimization problem. In a Bayesian framework, the solution of this
		optimization problem is interpreted as the maximum a posteriori
		estimator (see for example \cite{thiebaut2006introduction}). Introducing an
		objective function $\mathscr{C}$ to account for the instrumental effects, the
		statistical model of the measurements and regularization terms that promote
		more satisfying solutions (e.g., sparser and smoother solutions) and better
		reject noise,
		the problem writes:
		\begin{equation}
			\widehat{\V x} = \argmin_{\V x \geq \M 0}
			\; \mathscr{C}(\V r, \V x, \M A, \M \Omega, \V \mu)\,,
			\label{eq:global_argmin}
		\end{equation}
		 where:
		\begin{equation}
			 \mathscr{C}(\V r, \V x, \M A, \M \Omega, \V \mu) = \mathscr{D}(\V r, \M
			 A\,\V x,
			 \M \Omega) + \mathscr{R}(\V x, \V \mu)\,,
			\label{eq:global_argmin_d_r}
		\end{equation}
		and $\argmin_{\V x \geq \M 0}\; \mathscr{C}(\V r, \V x, \M A, \M \Omega, \V \mu)$
		denotes the set of nonnegative values of $\V x$ minimizing $\mathscr{C}(\V r,
		\V x, \M A, \M \Omega, \V \mu)$. The first term, $\mathscr{D}(\V r,
		\M A\,\V x, \M \Omega)$, is the data-fidelity term that penalizes the
		discrepancy between the recorded data $\V r$ and the modeled contribution $\M
		A\, \V x$ of the object of interest $\V x$ in the data. The data fidelity term
		also depends on parameters $\M \Omega$ that define the statistical
		distribution of the nuisance component $\V f$ (e.g., mean,
		variances or covariances, see Sect.~\ref{sec:learning_statistics}). The second
		term, $\mathscr{R}(\V x, \V \mu)$, is a regularization term enforcing some
		prior knowledge about the object properties to favor physically more plausible
		reconstructions and prevent noise amplification. The
		hyper-parameters $\V \mu$ tune the behavior of the regularization, in
		particular, its relative importance compared to the data fidelity term. In the
		following, we detail the ingredients of this general framework. The
		forward model $\M A$ of the off-axis PSF is developed in
		Sect.~\ref{sec:forward_model}. The statistical model of the nuisance
		component is stated in Sect.~\ref{sec:learning_statistics} while the
		estimation of the parameters $\M\Omega$ of this model is described in
		Sect.~\ref{sec:unbiased_estimation}.  Different possible regularizations are
		considered in Sect.~\ref{sec:regularized_inversion}. Finally, a strategy for
		the automatic tuning of the regularization hyper-parameters is presented
		in~\ref{sec:automatic_setting_regularization_parameters}.

		\subsection{Forward model: off-axis point spread function}
		\label{sec:forward_model}

		\begin{figure*}
			\centering%
			\includegraphics[width=\textwidth]{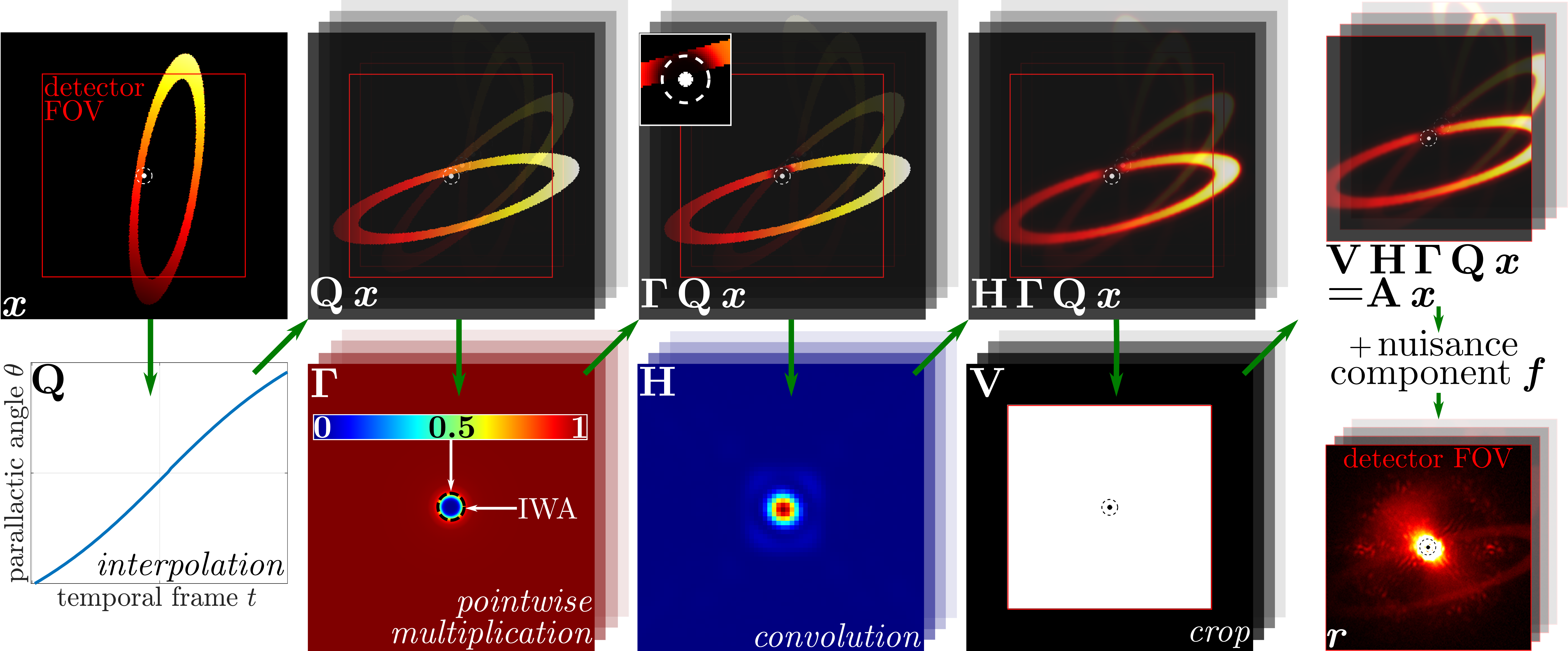}%
			\caption{%
				Schematic illustration of the forward model $\M A
				\, \V x = \M V \, \M H \, \M \Gamma \, \M Q \, \V x$ of \REXPACO.
				First, the object $\V x$ is rotated by the parallactic angles by operator
				$\M Q$. Then, each resulting temporal frame is attenuated by the (time
				independent) transmission of the coronagraph with $\M \Gamma$, and
				convolved
				by the off-axis PSF by application of $\M H$. Finally, the field of view
				seen by the sensor is extracted by $\M V$ to obtain the object
				contribution
				$\M A\, \V x$. The recorded ADI sequence is modeled by the sum of $\M A \,
				\V x$ and of the nuisance component $\V f$, as described by
				Eq.~(\ref{eq:additive_model}).}%
			\label{fig:principle}
		\end{figure*}

		The parameters $\V x \in \Reals_+^M$ represent the distribution of light due
		to the off-axis sources (e.g., circumstellar disks, exoplanets, brown
		dwarfs, background stars). During the acquisition of an ADI
		sequence, the field of view as seen by the detector
		rotates around the optical axis. After suitable instrument calibration, this
		apparent rotation is completely defined by the center of
		rotation and the parallactic angles recorded with the data\footnote{
		In practice for VLT/SPHERE-IRDIS, not all parallactic angles are measured and
		some of them are interpolated from others. Parallactic angles are also
		corrected for the estimated overheads of storage in a calibration and
		preprocessing step \citep{pavlov2008advanced}. Estimation errors of the
		rotation center and parallactic angles are not included in our statistical
		modeling.}. The contribution of $\V x$ in the $t$-th data frame $\V r_{t}$ is
		given by the linear model $\M A_{t}\,\V x$ where $\M
		A_{t}\from\Reals^{M}\to\Reals^{N}$ implements the off-axis point spread
		function for the considered frame: it models the apparent
		rotation of the field of view, the nonuniform attenuation by the coronagraph and the
		instrumental blurring. In other words, $[\M A\,\V x]_{t} \equiv \M
		A_{t}\,\V x$ for all frames $t \in \llbracket 1,T\rrbracket$ where $\M
		A\from\Reals^{M}\to\Reals^{N\times{}T}$ is the linear operator introduced in
		Sect.~\ref{sec:general_framework} and such that $\M A\,\V x$ models the
		contribution of $\V x$ to all data frames.

		Considering the size of the image reconstruction problem, it is essential that
		our model $\M A$ of the off-axis PSFs not only be an accurate approximation of
		the actual instrumental effects but also that $\M A$ and its adjoint be fast
		to apply in an iterative reconstruction algorithm (see
		Appendix~\ref{app:VMLMB}). To obtain a numerically efficient model,
		we decompose each operator $\M A_{t}$ into different factors: $\M Q_{t}$ to
		model the temporally and spatially variant effects due to the rotation of the
		field of view, $\M \Gamma$ to account for the attenuation close to the optical axis due
		to the coronagraph, and $\M H$ to implement the instrumental blurring
		(considered in first approximation as spatially and temporally invariant).
		Operators $\M Q_{t}$ can be implemented by interpolation operations while
		operator $\M H$ performs bi-dimensional convolutions that can be computed
		efficiently using fast Fourier transforms (FFTs). Modeling the instrumental
		blurring as being shift-invariant (a convolution) is accurate in most of the
		field of view except close to the optical axis. The coronagraph induces a reduction of
		the light transmission at short angular separations (i.e., under and near the
		coronagraphic mask) and a deformation of the PSF. We model the variable light
		transmission by a diagonal operator $\M \Gamma=\Diag(\V\gamma)$ with
		$\V\gamma$ a 2-D transmission mask. The mask $\V\gamma$ has a smooth profile
		with all entries in the range $[0,1]$; from 0 at the center of the coronagraph
		(i.e., no flux from off-axis sources transmitted) to 1 farther away, and being
		equal to 0.5 at the inner working angle (IWA) of the coronagraph.
		We neglect the deformation of the PSF in our model. This approximation has a
		limited effect since the area affected by the PSF deformations (close to or
		under the coronagraphic mask) is dominated by the nuisance component. An
		alternate way to handle the measurements in the area of the coronagraph would
		consist of assuming that they suffer an infinite variance (so that they be
		completely discarded in the subsequent estimations).
		Both approaches lead to	comparable reconstructions
		(i.e., differences in the reconstructed flux are several
		orders of magnitude lower than the disk magnitude) in all experimental
		cases we have considered.
		This model of the effect of the coronagraph on
		off-axis point sources as a simple attenuation varying with the angular
		position is consistent (for angular separations larger than $0.05$ arcsecond)
		with the calibrations of the coronagraphs considered in
		\cite{beuzit2019sphere} for the SPHERE instrument. For the results presented
		in Sect.~\ref{sec:results}, the experimental calibration of this transmission
		has been performed in the H-Johnson's spectral band (i.e., at $\simeq 1.6\,\micro\meter$) and extrapolated to the other
		wavelengths by using the experimental measurement and dedicated simulations,
		as described in \cite{beuzit2019sphere}.

		To summarize, our forward model of the contribution of the object of interest
		$\V x$ in the $t$-th frame writes\footnote{The forward model could be written
			differently so that the operations involved in $\M Q_{t}$, $\M \Gamma$ and $\M
			H$ would be applied in a different order, subject to slight adaptations.
			For example, if the field rotation and translation modeled by
				operator $\M Q_t$ are applied after the convolution $\M H$, the
				non-isotropic
				off-axis PSF has to be counter-rotated by the parallactic angles (i.e.,
				the
				blurring is not the same for all $t$) to counter-balance the effect of the
				permutation of $\M Q_t$ and $\M H$. The same remark applies for the
				attenuation $\M \Gamma$ if the mask $\V \gamma$ is not circular
				symmetrical.}:
		\begin{equation}
			\M A_{t} \, \V x = \M V \, \M H \, \M \Gamma \, \M Q_{t} \, \V x\,,
			\label{eq:forward_model}
		\end{equation}
		where $\M V$ is a simple
		truncation operator to extract the $N$ pixels corresponding to the actual data
		field of view from the (larger) $M$-pixels output of blurring operator $\M H$. Due to
		the apparent rotation of the field of view during the ADI sequence, previously unseen
		areas fall within the corners of the sensor and, by combining all
		measurements, a region larger than $N$ pixels can be reconstructed (up to 57\%
		more pixels).

		The factorization of the forward operator $\M A_t$ can straightforwardly be
		generalized to the case of a time-varying blur $\M H_t$ or transmission $\M
		\Gamma_t$ to account for an evolution of the observing conditions during the
		ADI sequence or a possible partial decentering of the coronagraph.
		Figure~\ref{fig:principle} illustrates the different steps to convert the
		object of interest $\V x$ into the model of its contribution into the ADI
		sequence recorded by the high contrast imaging instrument. In this
		figure, we adopt the same convention as for $\M A$ that, when the frame index
		$t$ of an operator is omitted, the result of applying the operator for all
		frames is considered.

		\subsection{Statistical model of the nuisance component}
		\label{sec:learning_statistics}

		The nuisance component $\V f$ in ADI sequences
		generally presents strong and nonstationary spatial correlations (speckles)
		that fluctuate over time, especially near the host star where the stellar
		leakages dominate, see Fig. \ref{fig:disk_data_fullfig}(a).
		Also, the presence of several bad pixels (hot, dead or randomly
		fluctuating) is a common issue in high contrast imaging where the images are
		recorded by infrared detectors. Despite a prereduction stage identifying and
		correcting for defective pixels, some of them displaying large fluctuations
		only on a few frames remain after this processing (see for example 
		\cite{pavlov2008advanced,delorme2017sphere} for the SPHERE instrument). We thus
		aim to account for the different causes of fluctuations of the
		nuisance component through a statistical model.

		We recover an image of the objects of interest $\V x$ based on a
		co-log-likelihood term $\mathscr{D}(\V r, \M A\,\V x, \M \Omega)$ that
		captures the statistical distribution of the residuals between $\V r$ and the
		model $\M A\,\V x$. Like in our previous works on the \PACO algorithm
		\citep{flasseur2018exoplanet, flasseur2018unsupervised, flasseur2018SPIE,
		flasseur2020pacoasdi, PACOrobuste}, we model the covariances locally at the
		scale of small patches of size $K$ (a few tens of pixels in practice).
		This approximation is justified by the nonstationarity of the spatial
		correlations and can be seen as a trade-off between considering full-size
		spatial covariance matrices (of size $N\times N$, and hence too large to be
		estimated and stored in practice) and completely neglecting the spatial
		correlations. Our choice amounts to model the nuisance component $\V f_t$
		of the $t$-th frame as the realization of independent Gaussian processes
		for each $K$-pixel patch, without overlap. Given that model, the probability
		density function of the nuisance component is given by:
		\begin{multline}
			\forall t,\quad\text{p}
			(\V{f}_t)\propto\prod_{n\in\mathbb{P}} \text{det}^{-\frac{1}{2}}(\M C_n) \, \\
			\times \exp\!\left[-{\textstyle\frac{1}{2}} \left(\M P_{n}(\V f_t-\V
			m)\right)\T \M C_n^{-1} \left(\M P_{n}(\V f_t-\V m)\right)\right]\,,
		\end{multline}
		with $\M P_{n}\from\mathbb{R}^{N} \to
		\mathbb{R}^{K}$ the operator extracting patch $n$ from a given temporal frame
		for any index $n$ in $\mathbb{P}$ defining the partition of the data frame
		into nonoverlapping $K$-pixel patches
        , and $\mathbb{P}$ is the list of the indices of the
        pixels at
        the center of the
        nonoverlapping patches.
        The vector $\V m\in \mathbb{R}^N$ denotes the
		temporal mean $\mathbb{E}_{t}(\V f_{t})$ while $\M C_n\in\mathbb{R}^{K\times
		K}$ denotes the temporal covariance matrix within the patch $n$.

		\begin{figure}
			\centering%
			\includegraphics[width=0.5\textwidth]{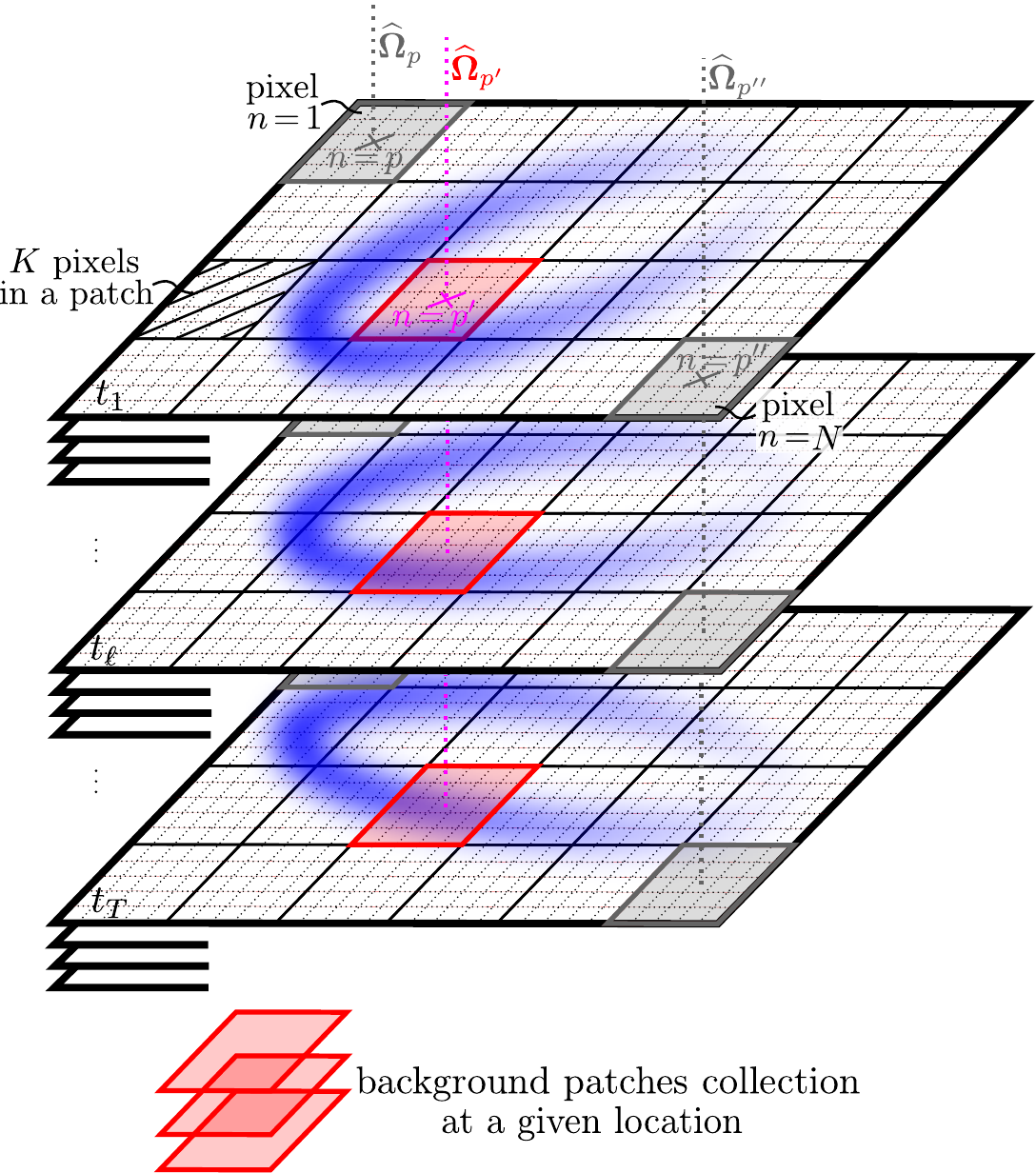}%
			\caption{%
				Extraction of patch collections for local learning of the statistics
				of the nuisance component from ADI sequences. The contribution of
				off-axis sources is shown in blue.}%
			\label{fig:rexpaco_principle}
		\end{figure}

		For $\V m$ and $\M C_{n}$, we use the same estimators that proved successful
		with \PACO for exoplanet detection: the temporal mean $\V m$ is estimated by
		the sample mean $\widehat{\V m}$ defined by:
		\begin{equation}
			\estim{\V m} = \frac{1}{T}\sum\limits_{t=1}^{T} \Paren{\V r_t - \M A_t\,\V
			x}\,,
			\label{eq:estimator_m}
		\end{equation}
		and the covariance matrices $\M C_{n}$ are estimated using a shrinkage estimator
		\citep{ledoit2004well,chen2010shrinkage}:
		\begin{equation} \ShrunkCov_n = \Paren*{1 -
			\widetilde{\rho}_n}\, \SampleCov_n + \widetilde{\rho}_n \, \widetilde{\M
			F}_n\,,
			\label{eq:shrinkage_convex_combination}
		\end{equation}
		that combines the high-variance and low-bias sample covariance estimator
		$\SampleCov_n$ with the low-variance and high-bias estimator $\widetilde{\M
		F}_n$ that neglects all covariances.  In other words, $\widetilde{\M F}_n$ is
		a diagonal matrix containing the pixel sample variances that are the diagonal
		entries of the sample covariance of the considered patch:
		\begin{equation}
			\SampleCov_n = \frac{1}{T} \sum_{t=1}^{T} \estim{\V u}_{n,t}\,\estim{\V
			u}_{n,t}\T \,,
			\label{eq:estimator_S}
		\end{equation}
		with:
		\begin{equation}
			\label{eq:residuals-def} \estim{\V u}_{n,t} = \M P_n\, \Paren[\big]{\V r_t -
			\estim{\V m} - \M A_t\,\V x}\,,
		\end{equation} the residuals in patch $n$ of
		frame $t$. With this specific choice for $\widetilde{\M F}_n$,
		Eq.~(\ref{eq:shrinkage_convex_combination}) rewrites:
		\begin{equation}
			\ShrunkCov_n = \widetilde{\M W}_n \odot \SampleCov_n\,,
			\label{eq:shrinkage_convex_combination_W}
		\end{equation}
		where $\odot$ stands for the Hadamard product (i.e., entrywise
		multiplication), and $\widetilde{\M W}_n \in \mathbb{R}^{K \times K}$ is
		defined for $\lbrace k, k'\rbrace \in \llbracket 1, K \rrbracket^2$ by:
		\begin{equation}
			\left[ \widetilde{\M W}_n \right]_{k,\,k'} =
			\begin{cases}
				1\ &\text{if } k =  k'\,, \\
				1-\widetilde{\rho}_n\ &\text{if } k \neq k'\,.
			\end{cases}
			\label{eq:W}
		\end{equation} %
		The shrinkage parameter $\widetilde{\rho}_n$ balances each estimator to reach
		a bias-variance trade-off. The extension of the results of
		\cite{chen2010shrinkage} to the diagonal covariance matrix $\widetilde{\M
		F}_n$ leads to the following data-driven expression for the shrinkage
		parameter $\widetilde{\rho}_n$ (see Eq.~(12) of
		\cite{flasseur2018exoplanet}):
		\begin{equation} \widetilde{\rho}_n =
			\frac{\Trace\Paren[\big]{\SampleCov_{n}^2} +
			\Trace^2\Paren[\big]{\SampleCov_{n}} -
			2\sum_{k=1}^K\Brack[\big]{\SampleCov{_{n}}}_{k,k}^2 }{ (T+1)\,\Paren[\Big]{
			\Trace\Paren[\big]{\SampleCov_{n}^2} -
			\sum_{k=1}^K\Brack[\big]{\SampleCov{_{n}}}_{k,k}^2 } } \,.
			\label{eq:shrinkage_rho}
		\end{equation} %
		It can be noted that most above defined quantities --- $\estim{\V m}$ in
		Eq.~\eqref{eq:estimator_m}, $\ShrunkCov_n$ in
		Eq.~\eqref{eq:shrinkage_convex_combination}, $\SampleCov_n$ in
		Eq.~\eqref{eq:estimator_S}, $\estim{\V u}_{n,t}$ in
		Eq.~\eqref{eq:residuals-def}, $\widetilde{\M W}_n$ in Eq.~\eqref{eq:W} and
		$\widetilde{\rho}_n$ in Eq.~\eqref{eq:shrinkage_rho} --- depend implicitly on
		the sought parameters $\V x$ although, and for the sake of simplicity, this is not
		always explicitly indicated. In Algorithms \ref{alg:update_stat} and
		\ref{alg:joint} presented in Sect.~\ref{sec:unbiased_estimation}, these
		quantities are labeled with the iteration number as the parameters $\V x$ change
		at each iteration.
		 Figure~\ref{fig:rexpaco_principle} illustrates the extraction of a
		collection of patches from an ADI sequence for locally learning
		the statistics of the nuisance component, in particular for the
		estimation of $\SampleCov_n$. The number $K$ of pixels in each patch, and
		hence the dimension of the local covariance matrices should be large enough to
		encompass the core of the off-axis PSF and yet not too large compared to the
		number $T$ of samples available for the estimation. The optimal patch size is
		estimated using the same criterion (validated empirically) as in \PACO and
		which yields an optimal patch size corresponding roughly to twice the off-axis
		PSF full width at half maximum (FWHM). This rule typically leads to
		$50 \lesssim K \lesssim 100$. Figure~\ref{fig:disk_data_fullfig}(c) gives
		examples of estimated patch covariance matrices $\ShrunkCov_n$ for
		different angular separations. It shows that the structure of the
		correlations strongly depends on the distance to the star and that neglecting
		these correlations would be a crude approximation, especially near the star
		were the nuisance component is highly correlated
		and fluctuating. Figure~\ref{fig:disk_data_fullfig}(c) also
		emphasizes that the covariances are well preserved by the regularization
		(i.e., $\widetilde{\rho}_n$ is small enough for $\SampleCov_n$ to represent
		the largest contribution). We have shown in
		\cite{flasseur2018exoplanet,PACOrobuste} that such a local model of the
		covariances approximates well the empirical distribution of the nuisance
		component. In \citet{PACOrobuste}, a small discrepancy was observed near the
		star due to the larger fluctuations\footnote{\samepage Large
			fluctuations in the nuisance component can be due to a slight decentering of
			the coronagraph or to a sudden degradation of the observing conditions. Finer
			statistical models of the nuisance component could be considered to account
			for these fluctuations. In particular, multivariate Gaussian scale mixture
			models \citep[GSM;][]{wainwright2000scale} have been shown to be very
			effective in the context of exoplanet detection
			\citep{flasseur2020pacoasdi,PACOrobuste}. Replacing
			the multi-variate Gaussian model consider in this paper by a GSM model
			requires specific developments that are left for future work.} of the nuisance
		component in this area. The impact of a statistical mis-modeling in this area
		is further discussed in Sect. \ref{sec:results_reconstruction_disk}.%

		Given the patch-based statistics $\widehat{\M \Omega} = \lbrace \widehat{\V
		m}, {\lbrace  \ShrunkCov_n \rbrace}_{n \in \mathbb{P}} \rbrace$ locally
		accounting for the fluctuations of the nuisance component, the
		co-log-likelihood rewrites:
		\begin{equation}
			\mathscr{D}(\V r, \M A\,\V x, \widehat{\M \Omega}) =
			\frac{1}{2} \sum\limits_{n \in \DisjointPatches} \Paren[\Big]{ \sum_{t=1}^{T}
			\Norm{\estim{\V u}_{n,t}}_{\ShrunkCov_n^{-1}}^2 + T\,\log \det{\ShrunkCov_n}
			} \,,
			\label{eq:patched_co-log-likelihood}
		\end{equation} %
		where the summation over $n$ is performed on the list $\DisjointPatches$ of
		nonoverlapping $K$-pixel patches while the residuals $\estim{\V
			u}_{n,t}$ in the patch $n$ for the frame $t$ are defined in
		Eq.~\eqref{eq:residuals-def}.

		\subsection{Unbiased estimation of the mean and spatial covariances}
		\label{sec:unbiased_estimation}

		\begin{algorithm}[t] 
			\SetAlgoVlined \DontPrintSemicolon\SetSideCommentLeft \caption{\REXPACO
				reconstruction\newline (Alternating estimation approach).}
			\label{alg:update_stat}%
			\hspace*{-2ex}%
			\begin{minipage}{0.98\columnwidth} \KwIn{ADI sequence $\V
					r$.}%
				\KwIn{Forward operator $\M A$.}%
				\KwIn{Regularization parameters $\V \mu$.}%
				\KwIn{Relative precision $\eta \in (0, 1)$.}%
				\smallskip \KwOut{Unbiased estimate $\estim{\V x}$ of the light
				distribution
					of off-axis sources.} 
				\smallskip 
				$\estim{\V x}^{[0]} \gets \M 0_M$ \; $i\gets 0$ \; 
				\Do{ $\Norm[\big]{\estim{\V x}^{[i]} - {\estim{\V x}}^{[i-1]}} >
					\eta\,\Norm[\big]{\estim{\V x}^{[i]}}$ }{ $i\gets i+1$\; 
					\medskip \AlgoStep{1}{Learn statistics of nuisance
					term.}
					$\estim{\V m}^{[i]} \leftarrow \frac{1}{T}\sum_{t=1}^{T} \Paren*{\V
					r_t -
						\M A_t\,\estim{\V x}^{[i-1]}}$
						\Comment*[r]{Eq.~\eqref{eq:estimator_m}}
					\For{$n \in \DisjointPatches$}{ \For{$t \in \llbracket
					1,T\rrbracket$}{ $\V
							u_{n,t}^{[i]} \gets \M P_{n} \Paren*{ \V r_t - \estim{\V
							m}^{[i]} - \M A_t
								\,\estim{\V x}^{[i-1]} }$ } $\SampleCov_n^{[i]}
								\leftarrow \frac{1}{T}
						\sum_{t=1}^{T} \V u_{n,t}^{[i]} \, \V u_{n,t}^{[i]}\,\T$
						\Comment*[r]{Eq.~\eqref{eq:estimator_S}}
						$\widetilde{\rho}_n^{[i]} \gets
						\frac{ \Trace\Paren[\big]{\SampleCov_{n}^{[i]\,2}} +
							\Trace^2\Paren[\big]{\SampleCov_{n}^{[i]}} -
							2\sum_{k=1}^K\Brack[\big]{\SampleCov_n^{[i]}}_{k,k}^2 }{
							(T+1)\,\Paren[\Big]{
							\Trace\Paren[\big]{\SampleCov_{n}^{[i]\,2}} -
								\sum_{k=1}^K\Brack[\big]{\SampleCov_n^{[i]}}_{k,k}^2 } }$
						\Comment*[r]{Eq.~\eqref{eq:shrinkage_rho}} \For{$(k,k') \in
						\llbracket
							1,K\rrbracket^2$}{ $\Brack[\big]{\widetilde{\M
							W}_n^{[i]}}_{k,k'} \gets
							\begin{cases} 1 &\text{if } k = k' \\ 1 - \widetilde{\rho}_n\
							&\text{if } k
							\neq k'\\ \end{cases}$ \Comment*[r]{Eq.~\eqref{eq:W}} }
							$\ShrunkCov_n^{[i]}
						\gets \widetilde{\M W}_n^{[i]} \odot \SampleCov_n^{[i]}$
						\Comment*[r]{Eq.~\eqref{eq:shrinkage_convex_combination_W}} }
						$\estim{\M
						\Omega}^{[i]} \gets \Brace{ \estim{\V m}^{[i]},
						\Brace{\ShrunkCov_n^{[i]}}_{n \in \DisjointPatches} }$\; \medskip
					\AlgoStep{2}{Reconstruct an image of the objects.} $\widehat{\V
					x}^{[i]}
					\gets \argmin\limits_{\V x \geq \M 0} \mathscr{D}(\V r, \M A\, \V x,
					\estim{\M \Omega}^{[i]}) + \mathscr{R}(\V x, \V \mu)$
					\Comment*[r]{Eq.~\eqref{eq:global_argmin}} } \end{minipage}
					\end{algorithm}

		\begin{algorithm}[h] 
			\SetAlgoVlined \DontPrintSemicolon\SetSideCommentLeft
			\caption{\REXPACO reconstruction\newline(Joint estimation approach).}
			\label{alg:joint}%
			%
			\hspace*{-2ex}%
			\begin{minipage}{0.98\columnwidth} \KwIn{ADI sequence $\V r$.} \KwIn{Forward
					operator $\M A$.} \KwIn{Regularization parameters $\V \mu$.}
					\smallskip
				\KwOut{Unbiased estimate $\estim{\V x}$ of the light distribution of
				off-axis
					sources.} 
				\smallskip 
				\AlgoStep{1}{Compute shrinkage matrices.} $\{\widetilde{\M
					W}_n\}_{n\in\mathbb{P}}$\; $\estim{\V m}^{[1]} =
				\frac{1}{T}\!\sum\limits_{t=1}^{T}\! \V r_t$
				\Comment*[r]{Eq.~\eqref{eq:sample_mean_1}} \For{$n \in \DisjointPatches$}{
					$\SampleCov_n^{[1]} \gets \tfrac{1}{T}\!\sum\limits_{t=1}^{T}\! \M
					P_{n}\!\Paren[\Big]{ \V r_t - \estim{\V m}^{[1]} } \!\Paren[\Big]{ \V
					r_t -
						\estim{\V m}^{[1]} }^{\hspace*{-0.75ex}\TransposeLetter}\M
						P_{n}\T$
					\Comment*[r]{Eq.~\eqref{eq:sample_covar_1}} $\widetilde{\rho}_n^{[1]}
					\gets
					\frac{ \Trace\Paren[\big]{\SampleCov_{n}^{[1]\,2}} +
						\Trace^2\Paren[\big]{\SampleCov_{n}^{[1]}} -
						2\sum_{k=1}^K\Brack[\big]{\SampleCov_n^{[1]}}_{k,k}^2 }{
						(T+1)\,\Paren[\Big]{ \Trace\Paren[\big]{\SampleCov_{n}^{[1]\,2}} -
							\sum_{k=1}^K\Brack[\big]{\SampleCov_n^{[1]}}_{k,k}^2 } }$
					\Comment*[r]{Eq.~\eqref{eq:shrinkage_rho}} \For{$(k,k') \in \llbracket
						1,K\rrbracket^2$}{ $\Brack[\big]{\widetilde{\M W}_n^{[1]}}_{k,k'}
						\gets
						\begin{cases} 1\ &\text{if } k = k' \\ 1-\widetilde{\rho}_n\
						&\text{if } k
						\neq k' \end{cases}$\Comment*[r]{Eq.~\eqref{eq:W}}} } \smallskip
				\AlgoStep{2}{Joint minimization with fixed $\{\widetilde{\M
						W}_n\}_{n\in\mathbb{P}}$.} $\widehat{\V x} \gets
						\argmin\limits_{\V x \geq
					\M 0} \mathscr{D}_{\text{joint}}(\V r, \V x) + \mathscr{R}(\V x, \V
					\mu)$
				\Comment*[r]{Eqs.~\eqref{eq:global_argmin} and
					\eqref{eq:co-log-likelihood_shrinkage}} \end{minipage} \end{algorithm}

		The estimators $\widehat{\V m}$, in Eq.~(\ref{eq:estimator_m}), and
		$\ShrunkCov_n$, in Eq.~(\ref{eq:shrinkage_convex_combination}), that appear in
		Eq.~(\ref{eq:patched_co-log-likelihood}) both depend on the object
		flux $\V x$ which is  unknown. To handle this problem, the simplest solution
		would be to neglect the object contribution as it is typically small
		compared to the nuisance component $\V f$. Under this
		assumption, Eqs.~(\ref{eq:estimator_m}) and
		(\ref{eq:estimator_S}) rewrite:
		\begin{subequations}
			\begin{align}
				\estim{\V m}^{[1]} &= \frac{1}{T}\sum\limits_{t=1}^{T} \V r_t \,,
				\label{eq:sample_mean_1} \\
				\SampleCov_n^{[1]} &= \frac{1}{T} \sum\limits_{t=1}^{T} \Paren*{\M P_{n}
				\Paren*{\V r_t - \estim{\V m}^{[1]}}}
				\Paren*{\M P_{n} \Paren*{\V	r_t - \estim{\V m}^{[1]}}}%
				^{\hspace*{-0.5ex}\TransposeLetter} \,,
				\label{eq:sample_covar_1}
			\end{align}
			\label{eq:estimator_sample_covs}%
		\end{subequations}%
		and correspond respectively to the sample mean and the sample covariances
		learned locally from the image patches.

		We show in Appendix~\ref{app:impcov} that neglecting the light flux of
		off-axis objects when characterizing the nuisance component leads to biased
		estimations, a problem common to many ADI post-processing methods and known as
		the self-subtraction problem 
		\citep{milli2012impact,pairet2019iterative}. To recover a better estimation of
		the object flux, it is then necessary to develop an unbiased estimation of the
		statistics $\M \Omega$. To do so, two strategies are possible: (i) an
		alternation between the reconstruction of $\estim{\V x}$ and the update of the
		statistics $\estim{\M \Omega}$ of the nuisance component $\V f$ from the
		residuals $\V r - \M A\,\estim{\V x}$, until convergence, as done in
		Algorithm~\ref{alg:update_stat}, or (ii) a joint estimation approach
		implemented by Algorithm~\ref{alg:joint}.  In this paper, we used $\eta =
		10^{-8}$ in the stopping condition of Algorithm~\ref{alg:update_stat} for all
		the reconstructions.
		Approach (ii) implemented by Algorithm~\ref{alg:joint} converges faster and is
		especially beneficial when reconstructing structures that are close to
		circular symmetry such as in
		Figs.~\ref{fig:hip80019_photometry_spiral_with_x_fullfig} and
		\ref{fig:hip80019_photometry_spiral_profiles_fullfig}. In order to be
		equivalent to Algorithm~\ref{alg:update_stat}, Algorithm~\ref{alg:joint}
		requires replacing the data-fidelity term $\mathscr{D}$ defined in
		Eq.~(\ref{eq:patched_co-log-likelihood}) by (the proof is detailed in Appendix~\ref{app:hierarchical}):
		\begin{multline}
			\mathscr{D}_{\text{joint}}(\V r, \V x) = \frac{T}{2}
			\sum\limits_{n \in \DisjointPatches} \log\det{\ShrunkCov_n}  \\
			+ \frac{1}{2} \sum\limits_{n \in \DisjointPatches}
			\Trace\Paren[\Big]{\ShrunkCov_n^{-1}\,\Paren[\Big]{ \widetilde{\M W}_n \odot
			\sum_{t=1}^{T} \widehat{\V u}_{n,t}\, \widehat{\V u}_{n,t}\T } }\,,
			\label{eq:co-log-likelihood_shrinkage}
		\end{multline}
		with $\estim{\V u}_{n,t}$ the residuals in the patch $n$ of frame $t$ defined in
		Eq.~\eqref{eq:residuals-def}. This modified
		data-fidelity corresponds to the co-log-likelihood under a Gaussian assumption
		except that shrinkage matrices $\widetilde{\M W}_n$ are
		introduced\footnote{Shrinkage matrices $\lbrace \widetilde{\M W}_n
			\rbrace_{n\in \mathbb{P}}$ are defined in Eq.~\eqref{eq:W} and we make the
			simplifying assumption that they are independent from $\V x$ so that they are
			estimated once for all given the data $\V r$.} so that the shrunk covariances
		$\ShrunkCov_n$ defined in Eq.~(\ref{eq:shrinkage_convex_combination_W}) are
		minimizers of $\mathscr{D}_{\text{joint}}$. The mean $\widehat{\V m}$ and
		shrunk covariances $\ShrunkCov_n$ can thus be replaced by their closed-form
		expressions given in Eqs.~(\ref{eq:estimator_m}) and
		(\ref{eq:shrinkage_convex_combination_W}) that depend on $\V x$ and the
		minimization be performed solely on $\V x$. Beyond the faster convergence of
		Algorithm~\ref{alg:joint}, the two algorithms differ slightly due to the
		shrinkage factors not being refined throughout the reconstruction in
		Algorithm~\ref{alg:joint}. We found the impact of that discrepancy to be
		barely noticeable in the reconstruction results and hence would recommend the
		general usage of Algorithm~\ref{alg:joint} for its improved speed. 

		\subsection{Regularization term}
		\label{sec:regularized_inversion}

		After discussing the data-fidelity term $\mathscr{D}(\V r, \M A\,\V x, \M
		\Omega)$ in the previous sections, we focus here on the regularization term
		$\mathscr{R}(\V x, \V \mu)$ and on the resolution of the inverse
		problem~(\ref{eq:global_argmin}).

		Regularization terms are introduced to enforce prior knowledge about the
		unknown object $\V x$ and to improve the conditioning of the inversion. Two
		classical regularizations are considered in the proposed method. The first one
		promotes smooth objects with sharp edges by favoring the sparsity of their
		spatial gradients. This is the goal of the so-called $\ell_2-\ell_1$
		edge-preserving regularization
		\citep{charbonnier1997deterministic,mugnier2004mistral,thiebaut2006introduction}
		widely used in image processing:
		\begin{equation}
			\mathscr{R}_{\text{smooth}}(\V x, \epsilon) = \sum\limits_{n=1}^N 
			\sqrt{ || \M \Delta_n \, \V x||_2^2+ \epsilon^2}
			\,,
			\label{eq:regularization_tv}
		\end{equation}
		where $\M \Delta_n$
		approximates the spatial gradient at pixel $n$ by finite differences and
		$\epsilon > 0$ sets the edge-preservation behavior of the
		regularization ($\epsilon = \sqrt{10^{-7}}$ in all the experiments of this paper): 
		Local differences $\M \Delta_n \, \V x$ that are below $\epsilon$
		in magnitude are smoothed similarly as a quadratic regularization would, while
		larger differences are preserved (e.g., border of a disk).
		This classical regularization is sufficiently flexible to remain adapted to
		different disk morphologies, as illustrated by 
		Figs.~\ref{fig:hip80019_photometry_ellipse_with_x_fullfig} to
		\ref{fig:hip80019_photometry_spiral_profiles_fullfig} where elliptical disks
		with sharp edges and spiral disks with smooth edges are considered.
		Since we aim to reconstruct a well-contrasted object on an dark background, we
		also consider a second regularization term promoting the sparsity of the
		object $\V x$ by penalizing its separable $\ell_1$-norm:
		\begin{equation}
			\mathscr{R}_{\ell_1}(\V x) = \sum\limits_{n=1}^N |x_n| \underset{\V x ≥ \V0}=
			\sum\limits_{n=1}^N x_n \,.
			\label{eq:regularization_l1}
		\end{equation}
		This sparsity-promoting regularization term is also well
		adapted to restore point-like sources. We discuss in
		Sect.~\ref{sec:results_joint} how an alternating estimation strategy can be
		applied to further improve the separation of overlapping point-like sources
		and extended structures. Due to the positivity constraint in
		Eq.~\eqref{eq:global_argmin} that guarantees that reconstructed light fluxes
		are nonnegative (i.e., $\V x ≥ \V0$), the $\ell_1$-norm in
		Eq.~\eqref{eq:regularization_l1} boils down to $\sum_n x_n$ which is a
		differentiable expression and smooth optimization techniques are applicable
		(see Appendix~\ref{app:VMLMB}). The two terms $\mathscr{R}_{\text{smooth}}$
		and $\mathscr{R}_{\ell_1}$ are combined by:
		\begin{equation}
			\mathscr{R}(\V x, \V \mu) =
			\mu_{\text{smooth}}\,\mathscr{R}_{\text{smooth}}(\V x, \epsilon) +
			\mu_{\ell_1}\,\mathscr{R}_{\ell_1}(\V x)\,,
		\end{equation} %
		where $\V \mu = \lbrace \mu_{\text{smooth}}, \mu_{\ell_1} \rbrace$ balances
		their relative weight with respect to the data-fidelity term $\mathscr{D}$.
		For a given set of parameters $\V \mu$, we solve the constrained
		minimization problem~(\ref{eq:global_argmin}) by running the VMLM-B algorithm
		\citep{thiebaut2002optimization}. Some technical elements about this minimization
		technique can be found in Appendix~\ref{app:VMLMB}.

		\section{Results}
		\label{sec:results}

		\subsection{Evaluation on simulated disks}

		\begin{figure*}[tp]
			\centering%
			\includegraphics[width=\textwidth]{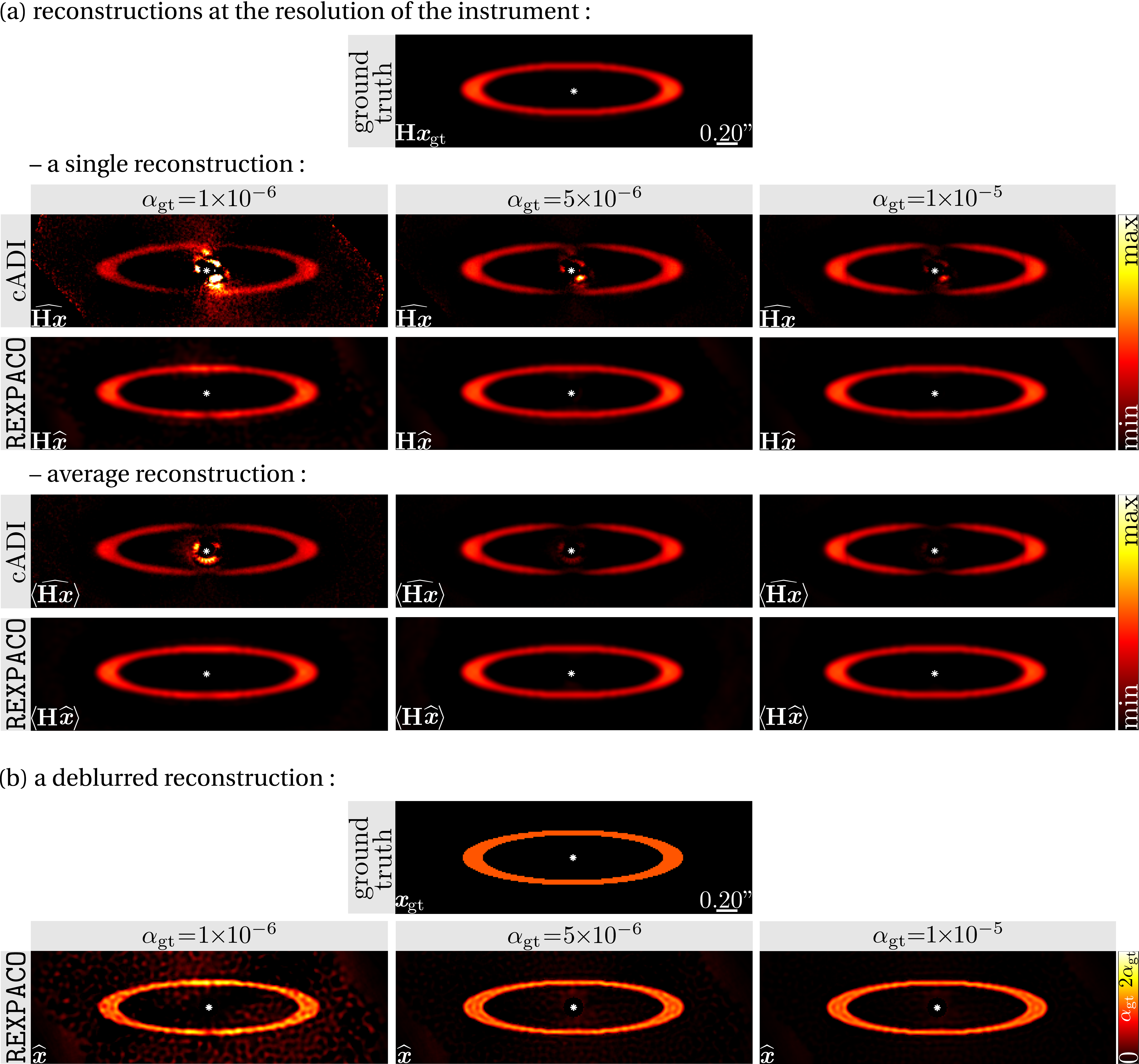}%
			\caption{%
				Reconstructions of a simulated elliptical disk: (a) comparisons between
				cADI
				image combinations and reblurred \REXPACO reconstructions; (b) resolution
				improvement achieved by deconvolution with \REXPACO. The average
				reconstructions are computed over ten injections of the simulated disk
				with
				various orientations with respect to the background.}%
			\label{fig:hip80019_photometry_ellipse_with_x_fullfig}
		\end{figure*}

		\begin{figure*}[tp]
			\centering%
			\includegraphics[width=\textwidth]{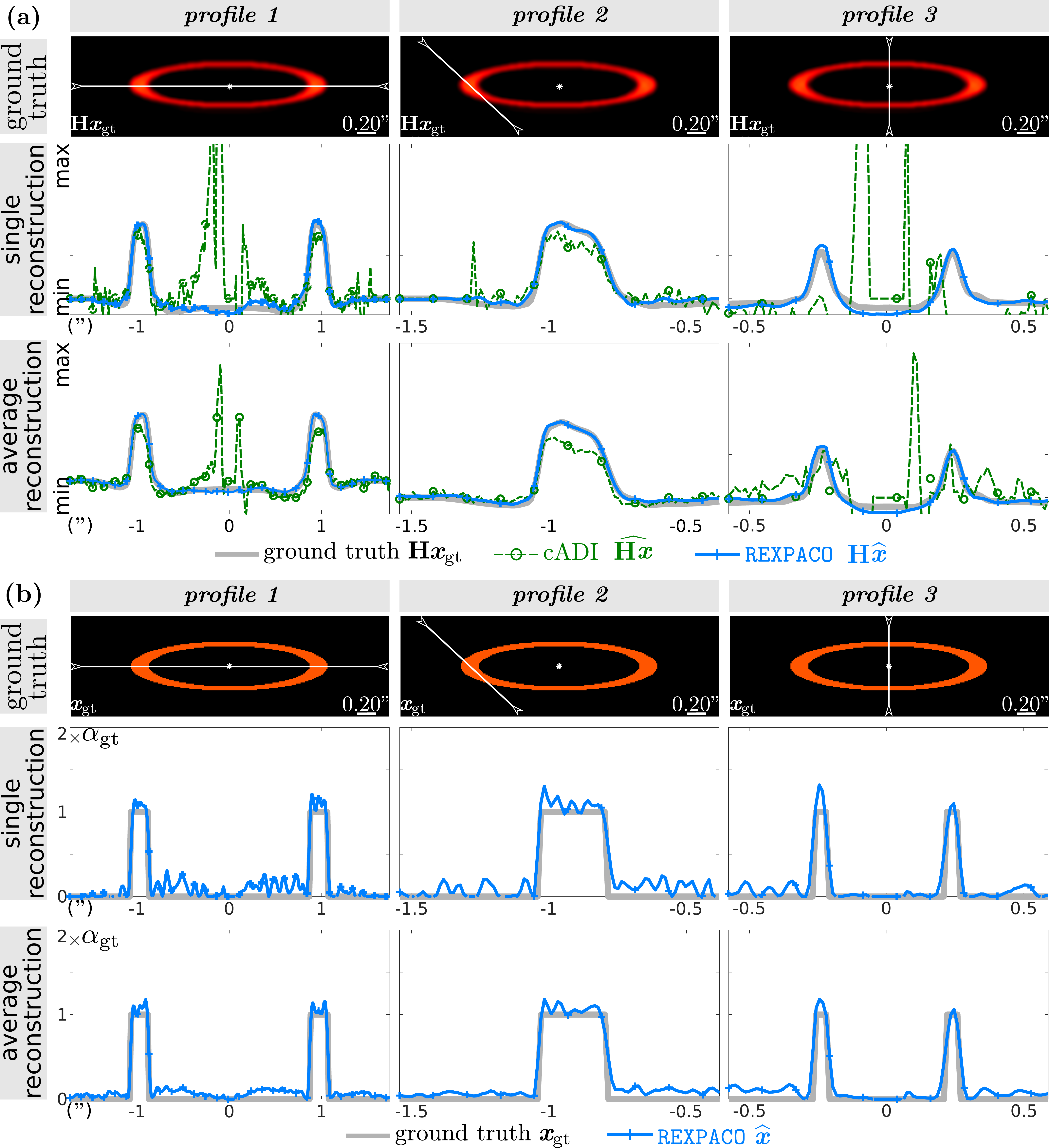}%
			\caption{%
				Line profiles extracted from the reconstructions at
				$\alpha_{\text{gt}}=10^{-6}$ shown in
				Fig.\ref{fig:hip80019_photometry_ellipse_with_x_fullfig}.}%
			\label{fig:hip80019_photometry_ellipse_profiles_fullfig}
		\end{figure*}

		\begin{figure*}[htbp]
			\centering%
			\includegraphics[width=\textwidth]{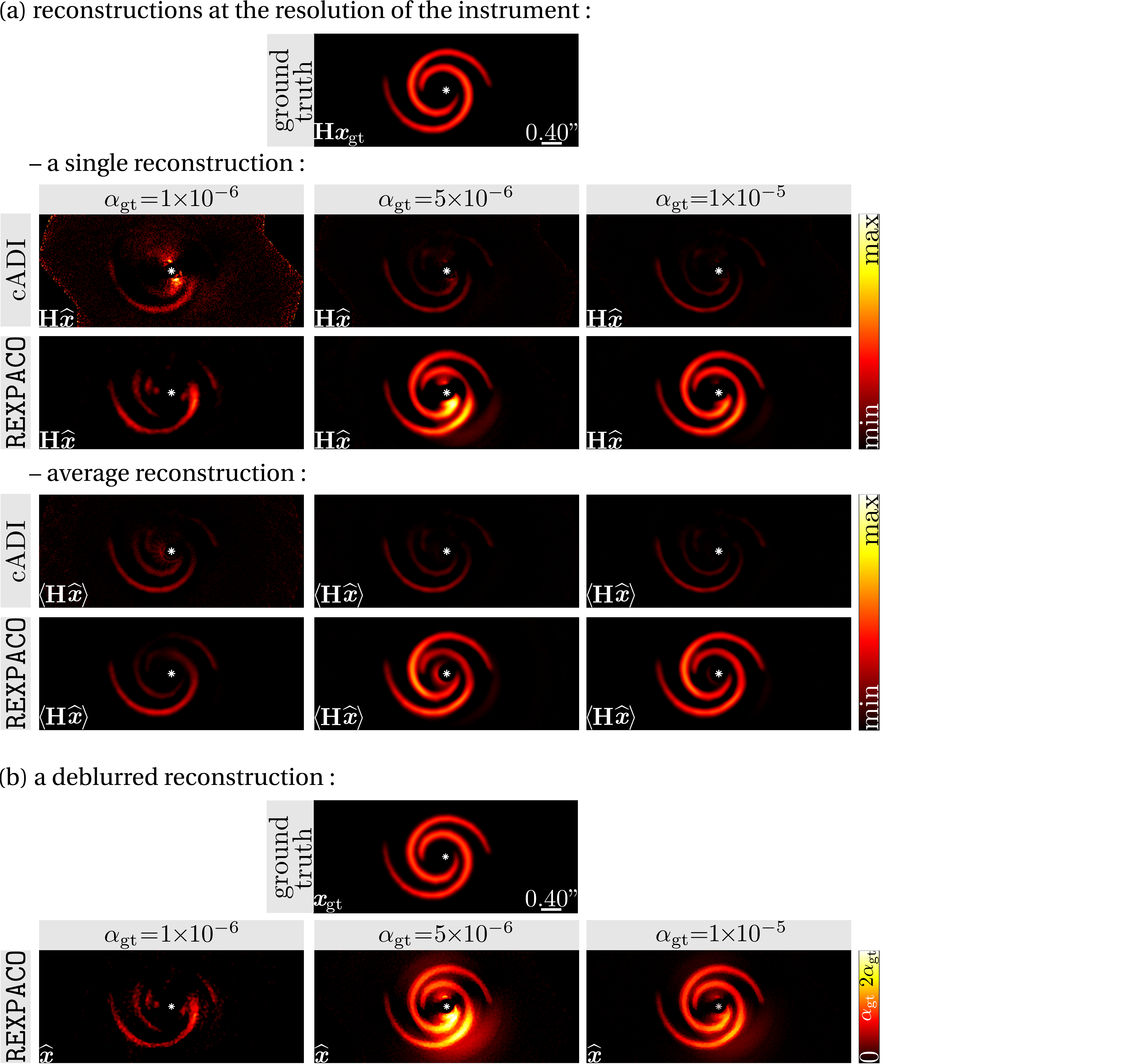}%
			\caption{%
				Reconstructions of a simulated spiral disk: (a) comparisons between cADI
				image combinations and reblurred \REXPACO reconstructions; (b) resolution
				improvement achieved by deconvolution with \REXPACO. The average
				reconstructions are computed over ten injections of the simulated disk
				with
				various orientations with respect to the background.}%
			\label{fig:hip80019_photometry_spiral_with_x_fullfig}
		\end{figure*}

		\begin{figure*}[t!] %
			\centering%
			\includegraphics[width=\textwidth]{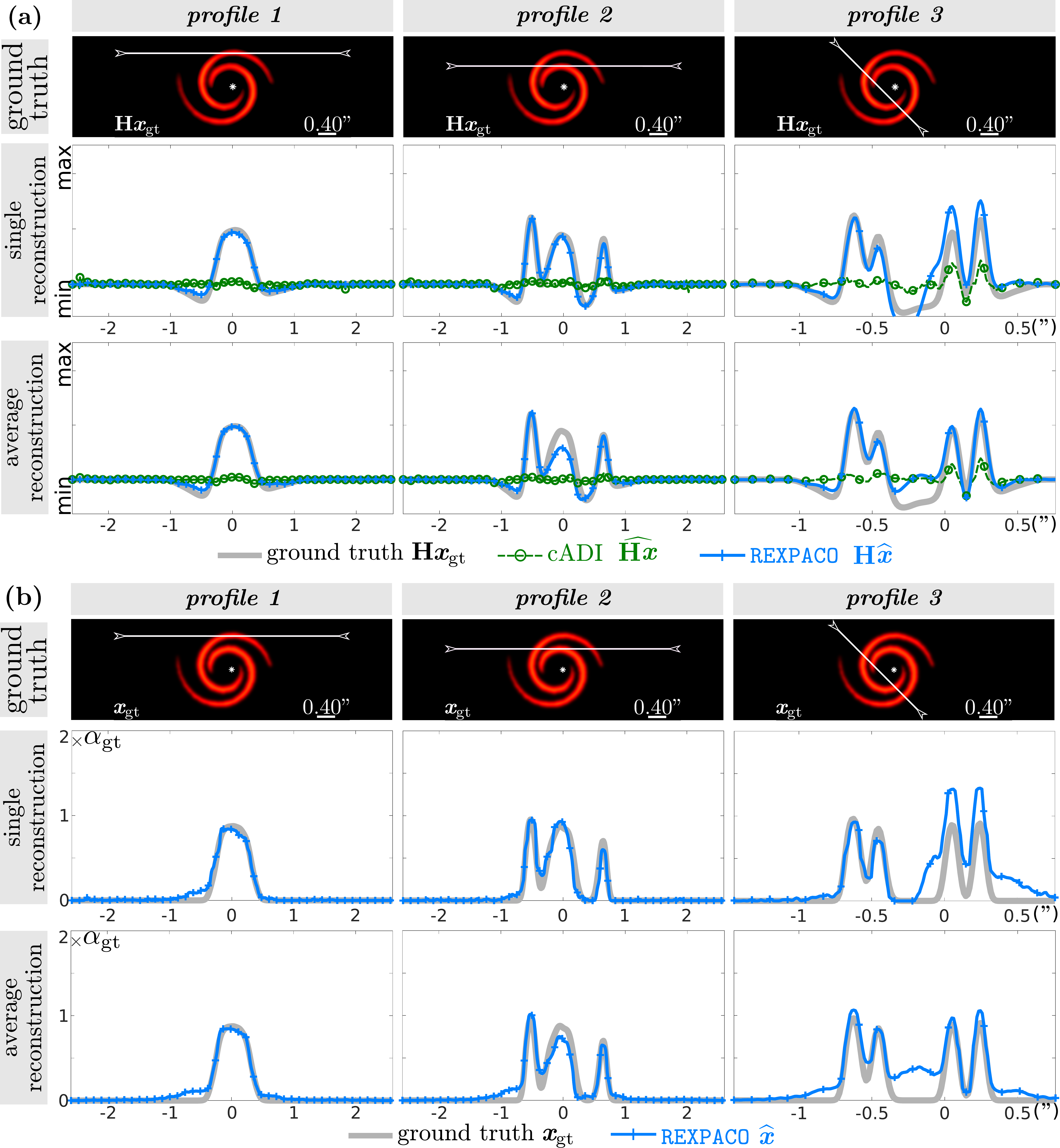}%
			\caption{ Line profiles extracted from the reconstructions at
				$\alpha_{\text{gt}}=10^{-5}$ shown in
				Fig.\ref{fig:hip80019_photometry_spiral_with_x_fullfig}.}
			\label{fig:hip80019_photometry_spiral_profiles_fullfig}
		\end{figure*}

		We first evaluate quantitatively, on numerical simulations, the ability
		of \REXPACO to estimate a faithful image of the light flux distribution of
		off-axis sources. We consider two simulated disks with very different
		morphologies: (i) a spatially centered elliptical disk with an eccentricity
		about 0.95 and with  sharp edges; (ii) a spiral disk with two arms whose
		center is shifted by ten pixels from the star center in one of two spatial
		directions. Contrary to the elliptical disk, the spiral disk has smooth edges.
		These simulated disks resemble the actual circumstellar disks presented in
		Sect.~\ref{sec:results_reconstruction_disk} so that our simulations can help
		to assess the quality of the reconstructions of real circumstellar disks. Each
		simulated disk is injected into a data set of HIP~80019 with no known off-axis
		source, at three different contrast levels: $\alpha_{\text{gt}} \in \lbrace
		1\times 10^{-6}, 5\times 10^{-6}, 1\times 10^{-5} \rbrace$. Reconstructions of
		the elliptical disk are performed with Algorithm~\ref{alg:update_stat} while
		the spiral disk that requires many more iterations of
		Algorithm~\ref{alg:update_stat} to converge has been processed with
		Algorithm~\ref{alg:joint}. A total of 60 reconstructions have been performed:
		for each disk and each level $\alpha_{\text{gt}}$, the simulated disk has been
		injected in one of ten different orientations with respect to the background in
		order to evaluate the mean and variance of the reconstructions.

		Simulations of the elliptical disk are reported in
		Figs.~\ref{fig:hip80019_photometry_ellipse_with_x_fullfig} and
		\ref{fig:hip80019_photometry_ellipse_profiles_fullfig}. We compare
		\REXPACO reconstructions to the cADI image combination technique. Since cADI
		does not perform a deconvolution, the comparison is performed at the
		resolution of the instrument, that is to say the images produced by cADI are shown next
		to the reblurred reconstructions $\M H\,\widehat{\V x}$ of \REXPACO in
		Figs.~\ref{fig:hip80019_photometry_ellipse_with_x_fullfig}(a) and
		\ref{fig:hip80019_photometry_ellipse_profiles_fullfig}(a). Large errors
		can be noted in cADI images, in particular at small angular separations. The
		computation of the average reconstruction obtained for the ten different
		orientations of the disk with respect to the background indicates the presence
		of systematic errors with cADI: an under-estimation of the light-flux in the
		region of the disk that is closest to the star (an issue due to
		limited angular diversity). In contrast, \REXPACO reconstructions are close to the
		ground truth, even for the lower level of contrast $\alpha_{\text{gt}} = 10^{-6}$.
		The deblurred reconstructions shown in
		Figs.~\ref{fig:hip80019_photometry_ellipse_with_x_fullfig}(b) and
		\ref{fig:hip80019_photometry_ellipse_profiles_fullfig}(b) are in good
		agreement with the ground truth. Unsurprisingly, the reconstruction quality is
		higher when the disk is brighter: more spurious fluctuations are visible in
		the deblurred reconstruction at $\alpha_{\text{gt}} = 10^{-6}$ than at
		$\alpha_{\text{gt}} = 5\times 10^{-6}$ or $\alpha_{\text{gt}} = 10^{-5}$.
		Although some discrepancies can be noted in the deblurred line profiles, the
		resolution is improved by the deconvolution process.

		Simulations of the spiral disk are reported in
		Figs.~\ref{fig:hip80019_photometry_spiral_with_x_fullfig} and
		\ref{fig:hip80019_photometry_spiral_profiles_fullfig}. The comparison
		with cADI shows a clear improvement of the reconstructions with \REXPACO: cADI
		fails to recover most of the disk. At the lowest contrast $\alpha_{\text{gt}}
		= 10^{-6}$ and owing to the near circular-symmetry of the structures close to
		the star, the spiral disk is challenging to disentangle from the nuisance
		component.  \REXPACO is however able to recover some parts of the spiral arms
		although their
		flux is underestimated. At the highest contrast $\alpha_{\text{gt}}
		= 10^{-5}$, line profiles of
		Fig.~\ref{fig:hip80019_photometry_spiral_profiles_fullfig} indicate that
		\REXPACO reconstructions are strongly improved compared to cADI and that some
		errors remain in particular at the lowest angular separations.

		The reconstructions are quantitatively evaluated by reporting the normalized
		root mean square error (N-RMSE, the lower the better) computed over different
		regions (the disk, the background, and the whole image): \begin{equation}
		\text{N-RMSE}(\V x_{\text{gt}}, \widehat{\V x}) = \frac{|| \V x_{\text{gt}} -
			\widehat{\V x}||_2}{||\V x_{\text{gt}}||_2}\,. \label{eq:n_rmse}
		\end{equation} Table~\ref{tab:nrmse_photometry} reports N-RMSE values and
		indicates a clear improvement over cADI: the errors are typically divided by a
		factor of three to four (the error reduction is more modest for the more challenging
		reconstructions of the spiral disk at the lowest fluxes).

		\begin{table*}[t!] %
			\caption{Quantitative evaluation of the reconstructions on simulated elliptical and spiral disks: N-RMSE, defined in
				Eq.~\eqref{eq:n_rmse}, for the reconstructions shown in
				Figs.~\ref{fig:hip80019_photometry_ellipse_with_x_fullfig} to
				\ref{fig:hip80019_photometry_spiral_profiles_fullfig} . The N-RMSE is
				also
				given
				on the restrictions $\mathcal{D}(\V x_{\text{gt}})$ and
				$\mathcal{D}(\widehat{\V x})$ to the area actually covered by the disks.
				The best scores are highlighted in bold fonts.}
			\centering%
			\begin{tabular}{ccccc} \toprule \midrule Score & Algorithm &
				$\alpha_{\text{gt}} = 1\times 10^{-6}$ & $\alpha_{\text{gt}} = 5\times
				10^{-6}$ & $\alpha_{\text{gt}} = 1\times 10^{-5}$ \\ \midrule & &
				\multicolumn{3}{c}{---\textit{ Elliptical disk, see
						Figs.~\ref{fig:hip80019_photometry_ellipse_with_x_fullfig} and
						\ref{fig:hip80019_photometry_ellipse_profiles_fullfig} }---} \\
				$\text{N-RMSE}\left(\mathcal{D}(\V x_{\text{gt}}),\mathcal{D}(\widehat{\V
					x})\right)$ & \REXPACO & 0.20 & 0.18 & 0.17 \\ $\text{N-RMSE}\left(\V
				x_{\text{gt}},\widehat{\V x}\right)$ & \REXPACO & 0.51 & 0.32 & 0.25
				\vspace{1mm} \\ $\text{N-RMSE}\left(\M H \, \V x_{\text{gt}},\M H \,
				\widehat{\V x}\right)$ & cADI & 0.65 & 0.45 & 0.43 \\
				$\text{N-RMSE}\left(\M
				H \, \V x_{\text{gt}},\M H \, \widehat{\V x}\right)$ & \REXPACO &
				\textbf{0.19} & \textbf{0.13} & \textbf{0.07} \vspace{2mm}\\ & &
				\multicolumn{3}{c}{---\textit{ Spiral disk, see
						Figs.~\ref{fig:hip80019_photometry_spiral_with_x_fullfig} and
						\ref{fig:hip80019_photometry_spiral_profiles_fullfig} }---} \\
				$\text{N-RMSE}\left(\mathcal{D}(\V x_{\text{gt}}),\mathcal{D}(\widehat{\V
					x})\right)$ & \REXPACO & 0.75 & 0.66 & 0.32 \\ $\text{N-RMSE}\left(\V
				x_{\text{gt}},\widehat{\V x} \right)$ & \REXPACO & 0.76 & 0.81 & 0.35
				\vspace{1mm} \\ $\text{N-RMSE}\left(\M H \, \V x_{\text{gt}},\M H \,
				\widehat{\V x}\right)$ & cADI & 0.88 & 0.87 & 0.89 \\
				$\text{N-RMSE}\left(\M
				H \, \V x_{\text{gt}},\M H \, \widehat{\V x}\right)$ & \REXPACO &
				\textbf{0.75} & \textbf{0.40} & \textbf{0.24} \\ \bottomrule
			\end{tabular}
			\label{tab:nrmse_photometry}
		\end{table*}

		\subsection{Automatic setting of the regularization parameters}
		\label{sec:automatic_setting_regularization_parameters}

		\begin{figure}[t!]
			\centering
			\includegraphics[width=0.5\textwidth]{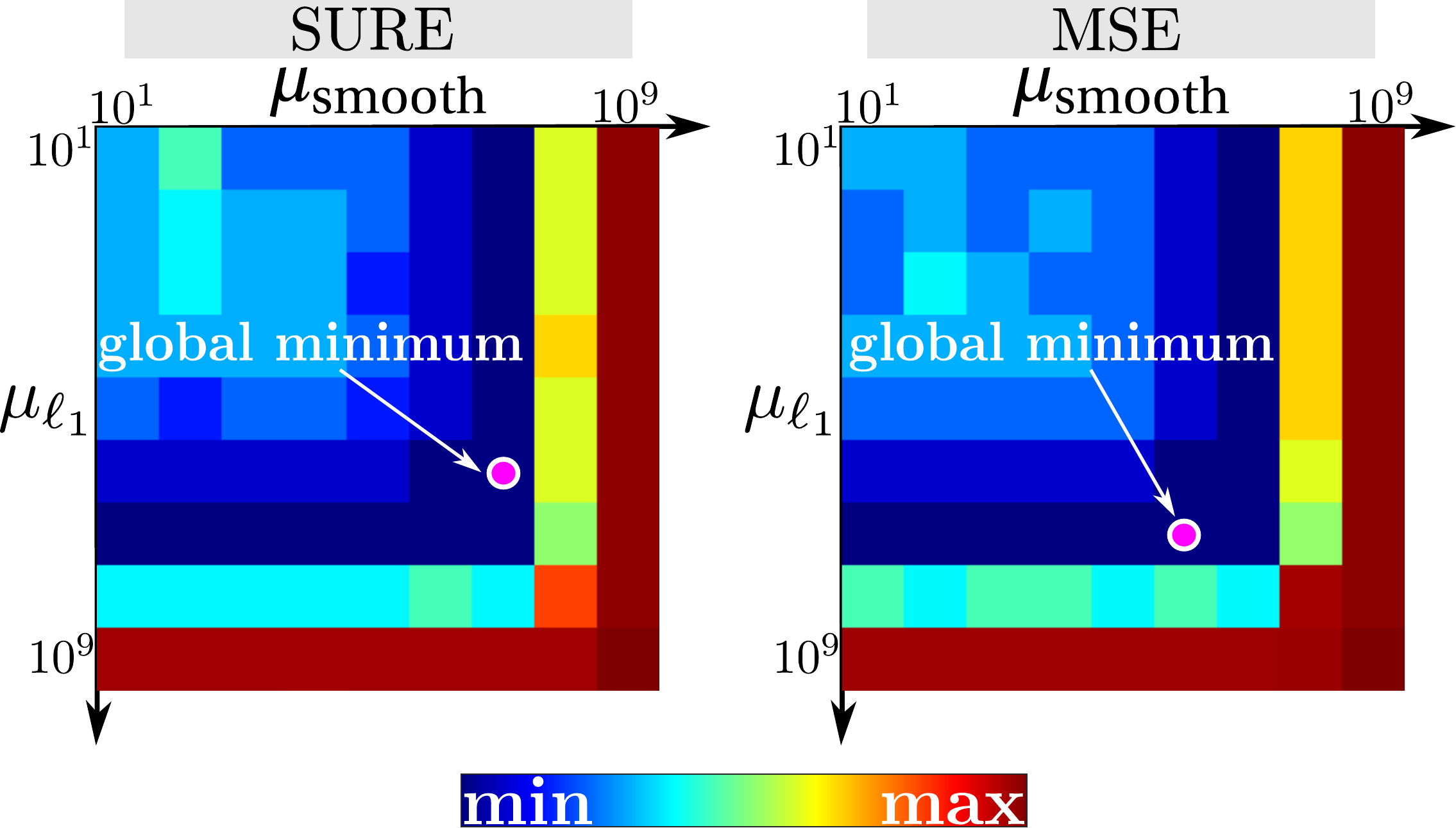}
			\caption{Comparison between the SURE
				criterion accounting for the fluctuations of the nuisance
					component through a local learning of the covariances
				(left,
					see Eq.~(\ref{eq:patched_sure_criterion})) and the true MSE
				(right, see Eq.~(\ref{eq:regularization_mse})).
				The pink circles represent the global minimum of each
					criterion.
					Data set: HIP 80019, see Sect.~\ref{sec:datasets_description} for
					observing
					conditions.}
			\label{fig:sure_criterion}
		\end{figure}

		\begin{figure*}[htbp] \centering
			\includegraphics[width=\textwidth]{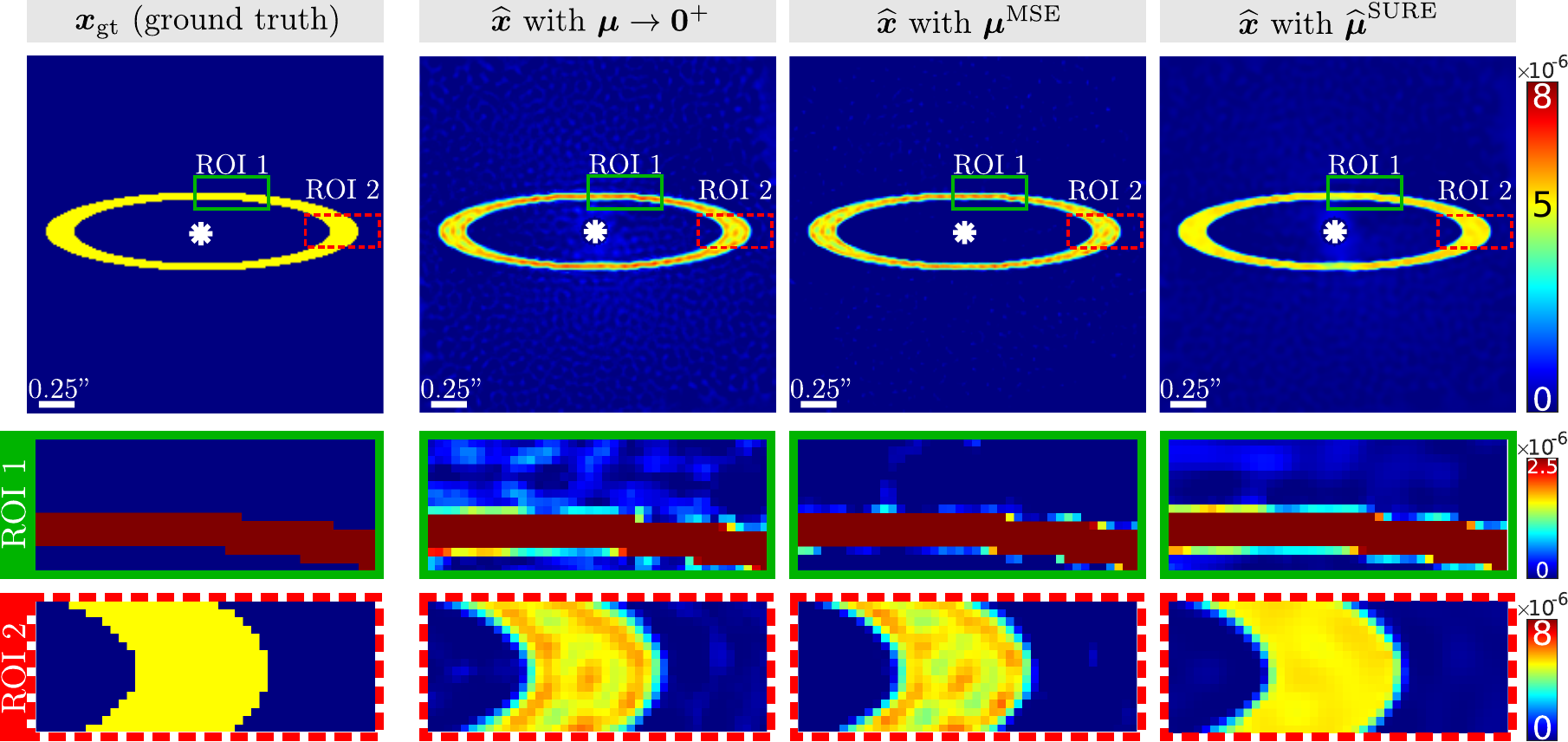}
			\caption{Influence of the
				regularization on the \REXPACO reconstructions. $\V
					x_{\text{gt}}$ (1\textsuperscript{st} column) is the ground truth
					elliptical
					disk injected at the level of flux $\alpha_{\text{gt}} = 1 \times
					10^{-5}$.
					The flux distribution $\estim{\V x}$ is reconstructed with a slight
					amount of
					regularization ($\V \mu \to \M 0^+$;  2\textsuperscript{nd} column),
					with $\V
					\mu^{\text{MSE}}$ derived by minimizing the optimal MSE criterion
					(3\textsuperscript{rd} column, see Eq.~(\ref{eq:regularization_mse}))
					and its
					estimate $\widehat{\V \mu}^{\text{SURE}}$ through the SURE criterion
					(4\textsuperscript{th} column, see
					Eq.~(\ref{eq:patched_sure_criterion})).
				The bottom panel shows zooms on two regions of interest (ROIs) extracted
				from
				the reconstructions given in the top panel. Data set: HIP
				80019,
					see Sect.~\ref{sec:datasets_description} for observing conditions.}
			\label{fig:sure_fullfig}
		\end{figure*}

		\begin{table}[t!] %
			\caption{RMSE scores of \REXPACO reconstructions of a simulated elliptical
			disk, for three different regularization levels. The best scores
			are highlighted in bold fonts.}%
			\centering%
			\begin{tabular}{cccc}
				\toprule \midrule
				Score $(\times 10^{-5})$ & $\V \mu \to \M 0^+$ & $\V
				\mu^{\text{MSE}}$ &
				$\widehat{\V \mu}^{\text{SURE}}$ \\
				\midrule
				RMSE (disk) & $4.37$ & $4.41$ & \bfseries{2.99} \\
				RMSE (background) & $5.19$ & \bfseries{2.54} & $3.96$ \\
				RMSE (whole field)& $6.79$ & \bfseries{4.96} & $5.09$ \\
				\bottomrule
			\end{tabular}
			\label{tab:rmse_regularization}
		\end{table}

		The setting of the parameters $\V \mu$ is of uttermost importance since the
		results of the regularized reconstructions depend both on the characteristics
		of the data set (e.g., observing conditions, amplitude of the
		field rotation) and on the properties of the objects to reconstruct
		(e.g., morphology, contrast). The parameters $\V \mu$ can be tuned manually by
		trial and error until the reconstruction is qualitatively acceptable, but this
		approach is not completely satisfactory since it is time-consuming and
		user-dependent.
		Several methods have been developed in the literature to automatically tune
		regularization parameters by minimizing a quantitative criterion
		\citep{craven1978smoothing,wahba1985comparison,stein1981estimation}. In this
		work, we consider the Stein's unbiased risk estimator approach
		\citep[SURE;][]{stein1981estimation}, which
		minimizes an estimate of the
		mean-square error (MSE) in the data space:
		\begin{equation}
			\text{MSE}(\V \mu) = ||\M A\left(\V x_{\text{gt}} -
			\estim{\V x}_{\V\mu}(\V r)\right)||_{\M{C}^{-1}}^2\,,
			\label{eq:regularization_mse}
		\end{equation}
		where  $\V x_{\text{gt}}$ is the (unknown)
		ground truth object, and $\estim{\V x}_{\V\mu}(\V r)$ denotes the
		reconstructed object $\widehat{\V x}$ obtained from the data $\V r$ when using
		regularization parameters $\V \mu$. For any given set of parameters $\V \mu$,
		the criterion $\text{SURE}(\V \mu)$ provides an unbiased estimation of
		$\text{MSE}(\V \mu)$ that does not require knowledge of the true object $\V
		x_{\text{gt}}$ \citep{stein1981estimation}:
		\begin{multline}
			\text{SURE}(\V \mu) = \Norm{ \V r - \estim{\V m} - \M A\,\estim{\V
			x}_{\V\mu}(\V r)}_{\M{C}^{-1}}^2 \\
			+ 2\Trace{\left(\M A \, \M J_{\V\mu}(\V r)\right)} - N\,T\,,
			\label{eq:sure_criterion}
		\end{multline}
		where $\M J_{\V\mu}(\V r) = \partial\estim{\V x}_{\V\mu}(\V
		r)/\partial\V r$ is the Jacobian matrix of the mapping $\V r\to\estim{\V
		x}_{\V\mu}(\V r)$ with
		respect to the components of the data $\V r$: $[\M J_{\V\mu}(\V
		r)]_{a,b}=\partial[\estim{\V x}_{\V\mu}(\V r)]_a/\partial[\V r]_b$. Since
		there is no closed-form expression for $\estim{\V x}_{\V\mu}(\V	r)$ (it is
		obtained by an iterative process), it is complex to derive $\Trace{\left(\M A
		\, \M J_{\V\mu}(\V r)\right)}$. An alternative proposed by
		\cite{girard1989fmcsure} considers a Monte Carlo perturbation method
		to numerically approximate this term by finite-differences:
		\begin{equation}
			\Trace{\left(\M A \, \M J_{\V\mu}(\V r)\right)} \approx \xi^{-1}\, \V b\T \,
			\M A \, \left[\estim{\V x}_{\V \mu}(\V r + \xi \V b) - \estim{\V x}_{\V\mu}(\V
			r) \right]\,,
			\label{eq:sure_approx_jacobian}
		\end{equation}
		where $\V b \in
		\mathbb{R}^{N\, T}$ is an independent and identically
			distributed pseudo-random vector of unit variance and $\xi$ is the amplitude
		of the perturbation. While the precise setting of $\xi$ is not a crucial point
		of the method, it must be chosen both large enough to prevent errors due to
		numerical underflows in the computation of the difference $\estim{\V
			x}_{\V\mu}(\V r + \xi \V b) - \estim{\V x}_{\V\mu}(\V r)$ and small enough so
		that the approximation in Eq.~(\ref{eq:sure_approx_jacobian})
		remains valid (the effects of the perturbation being nonlinear in the model).
		In practice, when we evaluate
		Eq.~(\ref{eq:sure_approx_jacobian}), we set $\xi = 0.1 \times
		\text{MAD}(\V r)$, where the median absolute deviation $\text{MAD}(\V
		r)=\text{median}(|\V r - \text{median}(\V r)|)$ is a robust estimator of the
		standard-deviation.

		As in our previous derivation of the statistical model of the nuisance
		component, we approximate the full covariance matrix $\M C$ that appears in
		Eq.~(\ref{eq:sure_criterion}) as block-diagonal, with blocks
		corresponding to the partition of the image into nonoverlapping patches. The
		patch covariance model and the perturbation approximation lead to the
		following risk estimator:
		\begin{multline}
			\text{SURE}(\V \mu)
			\approx \sum\limits_{n \in \DisjointPatches} \sum_{t=1}^{T} \Norm{\M P_n
			\left( \V r_t - \widehat{\V m} - \M A\,\estim{\V x}_{\V\mu}(\V r)
			\right)}_{\ShrunkCov_n^{-1}}^2\\
			+ (2/\xi) \, \V b\T \, \M A \, \left[ \estim{\V x}_{\V\mu}(\V r + \xi \V b) -
			\estim{\V x}_{\V\mu}(\V r) \right] - N\,T\,.
			\label{eq:patched_sure_criterion}
		\end{multline}
		The optimal value $\estim{\V \mu}^{\text{SURE}}$ of the
		regularization parameters is obtained by minimizing the
		quantity~(\ref{eq:patched_sure_criterion}) with respect to $\V \mu$.

		We have validated on a numerical example the automatic tuning of the
		regularization parameters $\V \mu$: The simulated elliptical disk is injected
		at the level of contrast $\alpha_{\text{gt}} = 10^{-5}$ on the data set of
		HIP~80019. Our SURE risk estimator given in
		Eq.~(\ref{eq:patched_sure_criterion}) is evaluated for
		different values of the parameters $\V \mu$. Figure~\ref{fig:sure_criterion}
		compares this criterion with the true MSE: The two criteria reach a global
		minimum for similar hyper-parameter values. The SURE criterion is
		intrinsically non-convex and additional non-convexities arise due to the
		sensitivity of the evaluation of the term $\Trace{\left(\M A \, \M
			J_{\V\mu}(\V r)\right)}$ by finite differences.  Rather than proceeding by
		local minimization of $\text{SURE}(\V\mu)$, it is therefore safer to
		systematically evaluate the criterion on a 2-D grid in order to identify the
		global minimum.

		Figure~\ref{fig:sure_fullfig} illustrates the impact of the regularization in
		three cases: an under-regularized setting (i.e., $\V
			\mu \to \M 0^+$), the automatic setting $\widehat{\V
			\mu}^{\text{SURE}}$ based on the minimization of the SURE
			criterion~(\ref{eq:patched_sure_criterion}), and the best-possible setting
		$\V \mu^{\text{MSE}}$ computed using the ground truth by
		minimizing the MSE criterion~(\ref{eq:regularization_mse}). The
		morphology of the reconstructed disk is improved by the regularization: It is
		smoother and sharper thanks to the edge-preserving term. The background of
		$\estim{\V x}$ is also less noisy (closer to zero) due to the $\ell_1$-norm
		penalization. It can be noted that in this simulation, the reconstructed
		object $\estim{\V x}$ with $\estim{\V \mu}^{\text{SURE}}$ is flatter and
		sharper than the disk reconstructed with the optimal parameters because the
		optimal parameters lead to a better reconstruction outside of the disk
		(reconstructed values are very close to zero: the ground truth value) at the
		expense of a slightly noisier reconstruction of the disk. In
		Figure~\ref{fig:sure_criterion}, the location of the global minimum (denoted
		by a circle) corresponds to parameters
		$\estim{\mu}_{\text{smooth}}^{\text{SURE}} > \mu_{\text{smooth}}^{\text{MSE}}$
		and $\estim{\mu}_{\ell_1}^{\text{SURE}} < \mu_{\ell_1}^{\text{MSE}}$, which
		leads to a stronger smoothing of the object with $\estim{\V\mu}^{\text{SURE}}$
		and a better rejection of close-to-zero noise in the background with
		$\estim{\V\mu}^{\text{MSE}}$. This qualitative observation is also
		confirmed by Table \ref{tab:rmse_regularization} which reports
		RMSE\footnote{Here, we use this metric instead of N-RMSE defined
				in Eq.~(\ref{eq:n_rmse}) since we aim to compare the absolute error made
				both
				inside and outside the spatial support of the disk.} scores between the
		ground truth $\V x_{\text{gt}}$ and the reconstructed flux distribution
		$\widehat{\V x}$ ($\text{RMSE} = \tfrac{1}{M}{||\V x_{\text{gt}} - \widehat{\V
				x}||_2}$). Table~\ref{tab:rmse_regularization} shows that the quality of
				the
		reconstruction obtained with the regularization parameters $\widehat{\V
			\mu}^{\text{SURE}}$ estimated by minimizing the SURE
		criterion~(\ref{eq:sure_approx_jacobian}) is very close to the best-possible
		reconstruction obtained by minimizing the MSE
		criterion~(\ref{eq:regularization_mse}).

		\subsection{Evaluation on data sets with circumstellar disks}

		In this section, we evaluate the ability of \REXPACO to reconstruct
		light flux distributions of actual circumstellar disks from ADI
		sequences. The results are compared to two existing methods,
		cADI and PCA (see Sect.~\ref{sec:introduction} for their
		respective principle), both designed for the detection of
		point-like sources but used extensively with different tuning parameters to
		process data with disks (see
		\cite{milli2012impact,pairet2019iterative} for studies
			of the resulting post-processing artifacts). We also compare \REXPACO to
		\PACO. The rationale behind this last comparison is to assess the benefit of
		the reconstruction framework of \REXPACO in addition to the common statistical
		modeling of the spatial covariances shared by the two methods. For cADI and
		the PCA, we used the SpeCal pipeline
		\citep{galicher2018astrometric} which is the post-processing standard of the
		SPHERE data center \citep{delorme2017sphere}.
		The \PACO implementation is based on the algorithm described in
		\cite{flasseur2018exoplanet,flasseur2018unsupervised,flasseur2018SPIE}. A
		\textsc{Matlab}\texttrademark ~implementation of the \REXPACO main routines is
		available
		online\footnote{\href{https://github.com/olivier-flasseur/rexpaco_demo.git}{\texttt{https://github.com/olivier-flasseur/rexpaco\_demo.git}}}
		 as a purpose of illustration. It is based on
		 \textsc{GlobalBioIm}\footnote{\url{https://biomedical-imaging-group.github.io/GlobalBioIm/}},
		 an opensource software for image reconstruction through an inverse problem
		framework \citep{soubies2019pocket}. Before post-processing with the tested
		algorithms, the data sets are preprocessed and calibrated with the prereduction
		tools \citep{pavlov2008advanced,maire2016sphere} of the SPHERE consortium available
		at the SPHERE Data Center \citep{delorme2017sphere}. In
		particular, background, flat-field, bad pixels, parallactic angles,
		wavelength, true-north and astrometric calibrations are performed during this step.

		\subsubsection*{Data sets description} \label{sec:datasets_description}

		For the comparisons, we have selected four data sets from the VLT/SPHERE-IRDIS
		instrument obtained by the observation of stars hosting known circumstellar
		disks of various morphological structures (e.g., face-on versus edge-on
		configuration, presence of inner and outer gaps, spirals, etc.).
		The rationale of such a selection is to test the versatility and
		the robustness of \REXPACO in reconstructing the flux distribution of
		circumstellar disks with various morphologies.
		The considered data sets were recorded
		on the following targets. Table~\ref{tab:dataset_logs} summarizes
		the main observing conditions of the corresponding
		ADI sequences.

				HR 4796A (HD 109573A) is the primary member of a
				binary system of the TW Hydrae association with an age about 12 Myr
				\citep{bell2015self} located at a distance about 72.8 pc
				\citep{van2007validation}. HR 4796A hosts a debris disk in
				face-on configuration first observed by the Hubble Space Telescope
				\citep{schneider1999nicmos}. Since then, its morphology has been studied
				intensively with ground-based facilities
				\citep{milli2017near,milli2019optical}. The relatively bright disk
				 (contrast between $10^{-5}$ and $10^{-4}$) exhibits a
				thin ring structure with a semimajor axis about 77 au, an inclination by
				about 76°, and a high surface brightness which suggest the presence of
				exoplanets although no companions have been detected to date.

				RY Lupi (HIP 78094) is a T-Tauri star of
				the Scorpius-Centaurus association with an age about 10-20 Myr
				\citep{pecaut2012revised} located at a distance about 151 pc
				\citep{brown2016gaia}. It hosts a transition disk directly
				observed for the first time in scattered light with the SPHERE instrument
				\citep{langlois2018first}. The disk is close to an edge-on
				configuration with an inclination about 70°, and it exhibits asymmetries
				and
				warped features near the star due to its interaction with the star
				magnetosphere. Observations in ADI and PDI have shown that
				the structure of the disk could be interpreted as spiral arms possibly due
				to the presence of putative low-mass exoplanets orbiting exterior to the
				spiral arms \citep{langlois2018first}.

				SAO 206462 (HD 135344B) is the secondary member of a
				binary system of the Upper Centaurus Lupus constellation with an age
				about 9
				Myr \citep{muller2011hd} located at a distance about 157 pc
				\citep{brown2016gaia}. SAO 206462 hosts a
				transition disk near a face-on configuration
				with an inclination about 11°, first resolved both in
				thermal
				emission \citep{doucet2006mid} and in scattered light
				\citep{grady2009revealing}. The disk exhibits two spiral
				arms, asymmetric features and an inner cavity. Its
				structure
				is of high interest since it could be due to the presence of low-mass
				exoplanets inside the arms or the cavity as shown by recent
				analysis of high contrast and high resolution observations
				\citep{maire2017testing}.

				PDS 70 (V1032 Centauri) is a T-Tauri star
			of the Scorpius-Centaurus association with an age about 5 Myr
			\citep{muller2018orbital} located at a distance about 113 pc
			\citep{brown2016gaia}. It hosts a protoplanetary disk and two
			confirmed exoplanets (PDS 70b and PDS 70c) in formation inside
			the disk. The protoplanetary disk hosted by PDS 70 was first
			resolved with the Naos Conica instrument (NaCo; \cite{lenzen2003naos}) at
			the VLT \citep{riaud2006coronagraphic}. It exhibits complex structures in
			the form of several arcs, outer and inner gaps, and potential spiral arms at
			the north side of the outer disk
			\citep{keppler2018discovery,mesa2019vlt}. The
			exoplanet PDS 70b has been detected by direct
			imaging with the VLT/SPHERE instrument \citep{keppler2018discovery} in near
			infrared, and the exoplanet PDS 70c has been
			unveiled with the VLT/MUSE instrument in $\text{H}_\alpha$
			\citep{haffert2019two}. To date, this is the unique system
			embedding multiple nascent exoplanets. The analysis of the structure of this
			system is extremely valuable to understand exoplanet formation mechanisms
			and raises still-open questions like the presence of additional putative
			exoplanets \citep{mesa2019vlt}.

		\begin{table*}[t!] \begin{threeparttable} \centering \caption{
			Observing conditions of ADI sequences from the VLT/SPHERE-IRDIS
			instrument
			considered in this paper.}%
		\begin{tiny} \begin{tabular}{cccccccccccc} \toprule \midrule Target & ESO
				ID
				& Obs. date & $T$ & $\lambda$ ($\micro\meter$) &
				$\Delta_{\text{par}}$ (°) & NDIT &
				DIT (s) &
				$\tau_0$ (ms) & Seeing ('') &
				Related paper\\
				\midrule HR 4796A & 095.C-0298 & 2015-02-03 & 110 &
				1.59 &
				48.5 & 8 & 32 &
				11.8
				& 0.67 &
				\cite{milli2017near}\\ RY Lupi & 097.C-0865 & 2016-04-16 & 80 &
				1.59 & 71.0 & 4 & 64
				& 61.4 &
				0.36 & \cite{langlois2018first}\\ SAO 206462 & 095.C-0298 &
				2015-05-15 & 63
				& 2.11 & 63.6 & 4 &
				64 & 85.0 &
				0.50 & \cite{maire2017testing}\\ PDS 70 & 100.C-0481 & 2018-02-24
				& 90 &
				2.11 & 95.7 & 3 & 96
				& 71.5 &
				0.46 & \cite{muller2018orbital}\\ HIP~80019$^{(*)}$ & 095.C-0389 &
				2015-04-21 & 116 & 2.11 & 42.8 & 11
				&
				8 & 23.2 & 0.80 &
				-\\ \bottomrule
		\end{tabular} \end{tiny} \begin{tablenotes} \item  \hspace{-5mm}
			\textbf{Notes.} Columns are: target name, ESO survey ID, observation
			date,
			number $T$ of available temporal frames, spectral filter
				$\lambda$, total amount of field rotation
			$\Delta_{\text{par}}$ of
			the field of view, number NDIT of sub-integration
				exposures,
				individual exposure time DIT, DIMM coherence time $\tau_0$,
			seeing value at
			the beginning of the observations and the first paper
				reporting analysis of the same data. All the
				observations are
				performed with the apodized Lyot coronagraph (APLC;
				\cite{carbillet2011apodized}) of the VLT/SPHERE instrument.
			$^{(*)}$ The
			ADI sequence of HIP 80019 was made of 176 temporal
			frames but
			we have discarded the 60 first frames manually. This
				sequence
				contains no detectable off-axis sources based on the application
				of \PACO
				and \REXPACO algorithms. If off-axis sources are present at an 
				undetectable level of contrast, they would introduce a negligible bias in
				comparison to
				the levels of contrast considered in our experiments. This ADI
				sequence is
			used in Sect.~\ref{sec:principle} to perform
			injections of
			fake disks.
		\end{tablenotes}
		\label{tab:dataset_logs}
		\end{threeparttable}
		\end{table*}

		\subsubsection*{Reconstruction of circumstellar disks}
		\label{sec:results_reconstruction_disk}

		We apply \REXPACO to the four selected ADI sequences presented
		in Table~\ref{tab:dataset_logs}. For each, the
		computation takes a few hours (\textsc{Matlab}\texttrademark ~implementation,
		processor Intel i9-9880H at 2.3 GHz). Figure~\ref{fig:reconstructions_fullfig}
		displays the results obtained with \PACO, PCA and cADI. For the
		PCA, we have tested several choices for the numbers of modes
		used in the decomposition and selected the best choice,
		corresponding to a small number of modes in order to limit the issue of
		self-subtraction (see \cite{pairet2019iterative} for a
			discussion about this effect). The images produced by \PACO are
		obtained with a maximum likelihood estimator of the light flux derived for
		point-like sources.
		For \REXPACO, the results take the form of reconstructed flux
		distributions $\estim{\V x}$. We also give the blurred versions
		$\M H \, \estim{\V x}$ to simplify the comparison with methods that do not
		perform a deconvolution. For the PCA and cADI, the results take
		the form of residual images obtained after subtraction of the estimated
		on-axis PSF. These residual images are normalized to the star flux
		that is encoded in the on-axis PSF provided with the data.
		Thus, the resulting images can also be interpreted as
		blurred flux distributions of the off-axis sources.

		Figure~\ref{fig:reconstructions_fullfig} shows that the mean levels of flux
		estimated by PCA and cADI differ significantly from the flux
		estimated by \REXPACO and \PACO\footnote{Fluxes estimated by \REXPACO and
		\PACO are approximately comparable. The residual flux
		difference is due to the fact that \PACO is based on a point-like source
		assumption which can lead to bias in the estimated flux of extended
		features.}: The PCA and cADI are sensitive to several causes of
		artifacts severely altering both the morphology and the
		photometry of the disks. In particular, cADI and the PCA
		suffer from positive and negative replicas and also from the
		self-subtraction phenomenon. cADI performs much better in the presence of
		extended features like disks due to its limited attenuation of
		the disk signal. However, cADI is also sensitive to the ADI artifacts
		creating morphological distortions like nonphysical discontinuities in the
		structure of the disks, as discussed in Sect.~\ref{sec:introduction}. 
		This is especially the case for the thin and bright
		ring surrounding HR 4796A. As discussed in
		Sect.~\ref{sec:unbiased_estimation}, the \REXPACO reconstruction exhibits an
		almost continuous elliptical structure, a flux asymmetry on the west side of
		the ring as predicted by intensity scattering models \citep{milli2017near},
		and the scattering structures predicted by radiative transfer models near the
		disk extremities \citep{lagrange2012insight}. The morphological structure of
		the reconstructed disk is very close the one obtained from polarimetric
		observations that are almost exempt from artifacts at such separations
		\citep{milli2019optical}. The faithfulness of the \REXPACO reconstruction is
		also supported by the simulations presented in
		Sect.~\ref{sec:unbiased_estimation} (see
		Figs.~\ref{fig:hip80019_photometry_ellipse_with_x_fullfig} and
		\ref{fig:hip80019_photometry_ellipse_profiles_fullfig}) where we injected a
		simulated elliptical disk with comparable eccentricity and spatial extent than
		the HR 4796A disk. In this latter case, no significant artifacts were
		identified in the reconstruction even though the simulated disk was one order
		of magnitude fainter than the HR 4796A disk.  The \REXPACO reconstruction of
		RY Lupi also exhibits reduced ADI artifacts. As seen on
		Fig.~\ref{fig:reconstructions_fullfig}, the \REXPACO reconstructions show
		better the smooth and faint extended disk structures. In addition, the
		reconstructed flux distribution uncovers new details around the brightest part
		of the disk such as spirals as well as the bottom and top disk planes.

		\begin{figure*}[htbp]
			\vspace{-2mm}
			\centering
			\includegraphics[width=\textwidth]{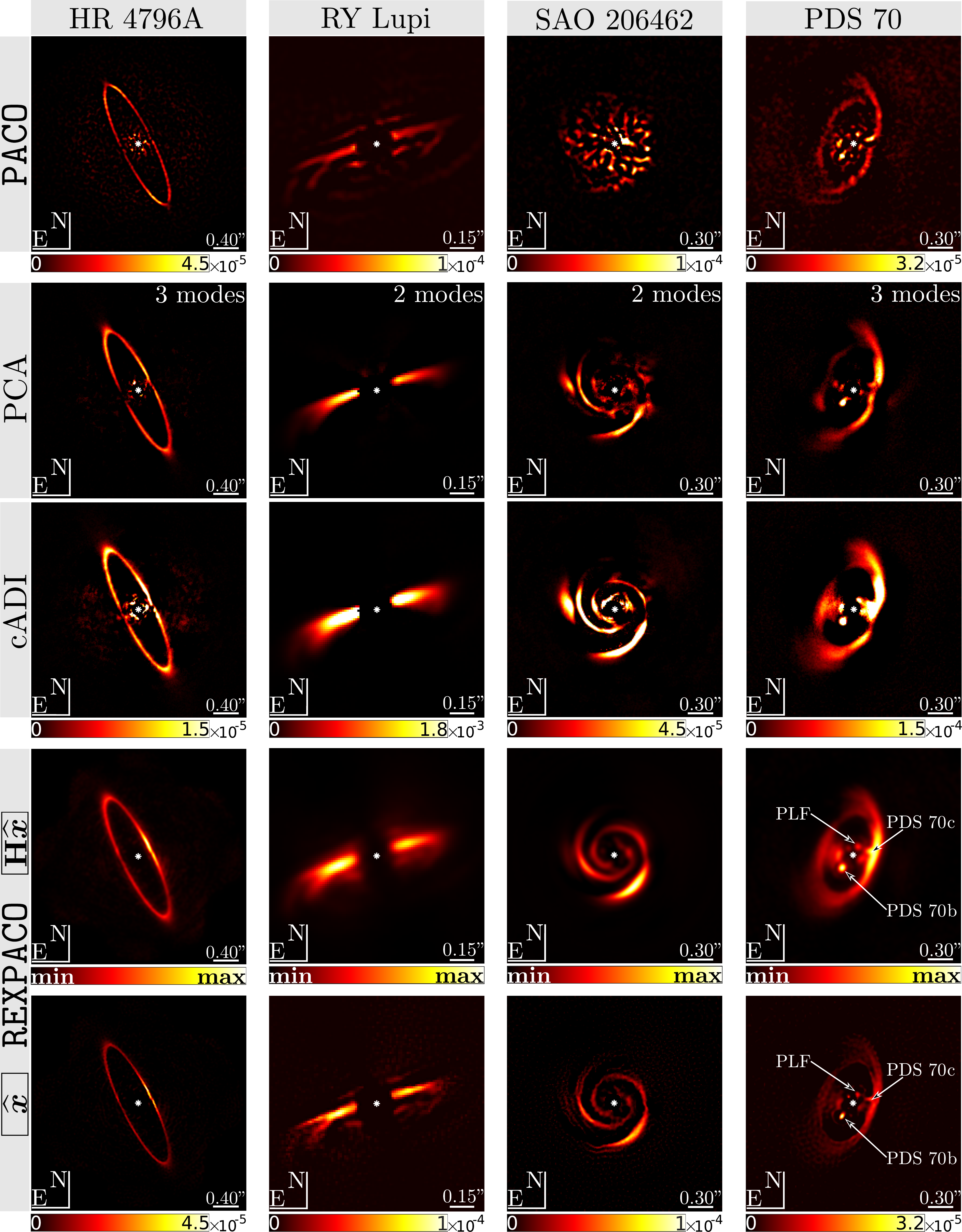}
			\caption{Images of the flux distribution reconstructed by
			\REXPACO comparatively to the \PACO, PCA and cADI algorithms.
				For \REXPACO, both the deblurred reconstruction
				$\widehat{\V x}$ and its reblurred version $\M H \, \widehat{\V x}$ are
				given for comparison with the other considered methods that do not
				produce deblurred images of the flux distribution.
				Data sets: HR 4796A, RY Lupi, SAO 206462 and PDS
				70, see Sect.~\ref{sec:datasets_description} for observing
					conditions.}
			\label{fig:reconstructions_fullfig}
		\end{figure*} %

		\begin{figure*}[tbp] \centering%
			\includegraphics[width=\textwidth]{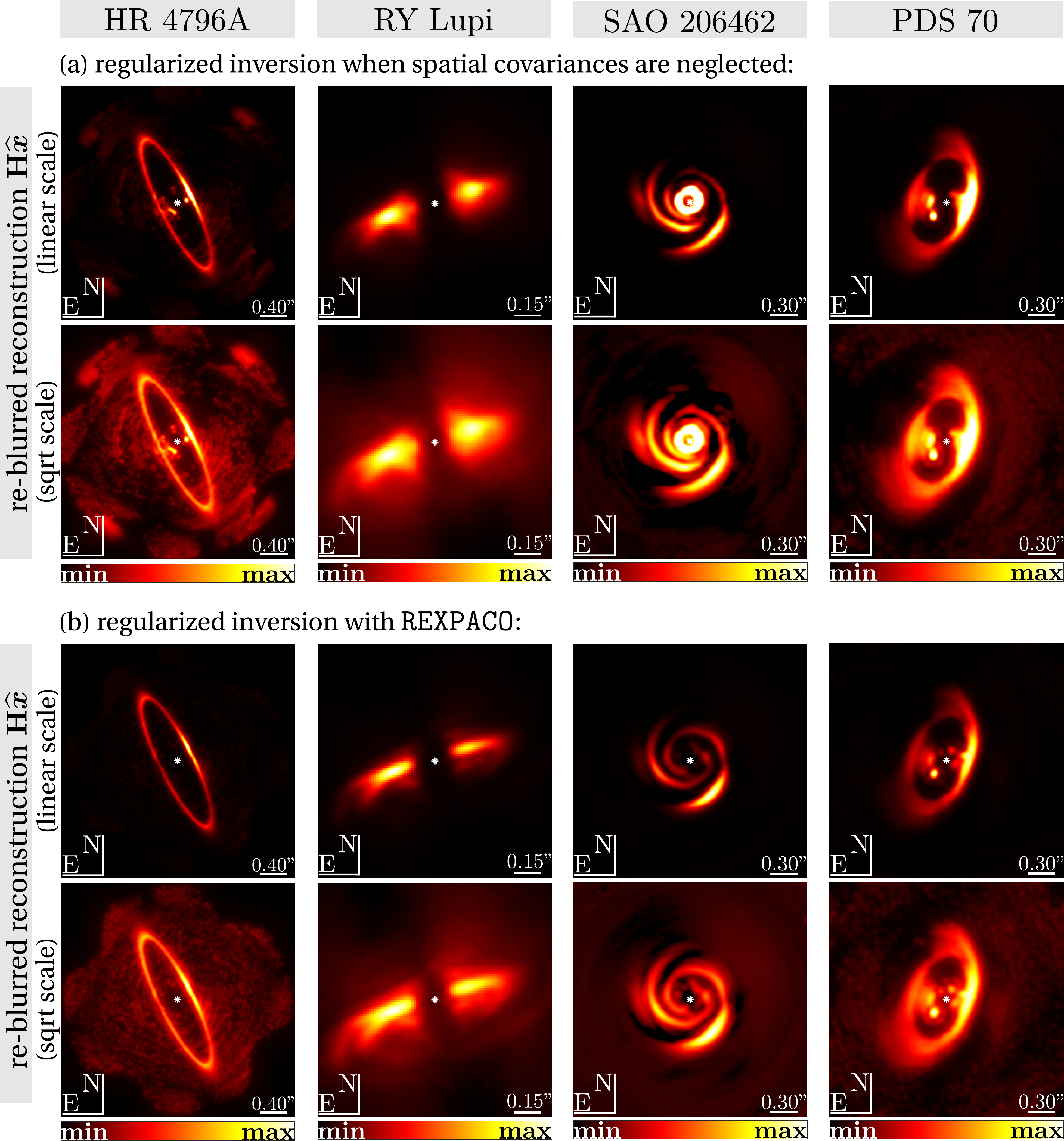}%
			\caption{%
				Importance of the modeling of spatial covariances. Comparison between (a)
				regularized inversions with a model of the nuisance component that
				accounts for the nonstationary variances but neglect covariances, and
				(b) \REXPACO inversions. Reconstructions are given first with a linear
				scale, then with a square-root scaling in order to better distinguish the
				lowest reconstructed values. Data sets: HR 4796A, RY Lupi, SAO 206462 and
				PDS 70, see Sect.~\ref{sec:datasets_description} for observing
				conditions.}%
			\label{fig:reconstruction_diagcov_fullfig}
		\end{figure*}

		\begin{figure}[t!]
			\centering
			\includegraphics[width=0.5\textwidth]{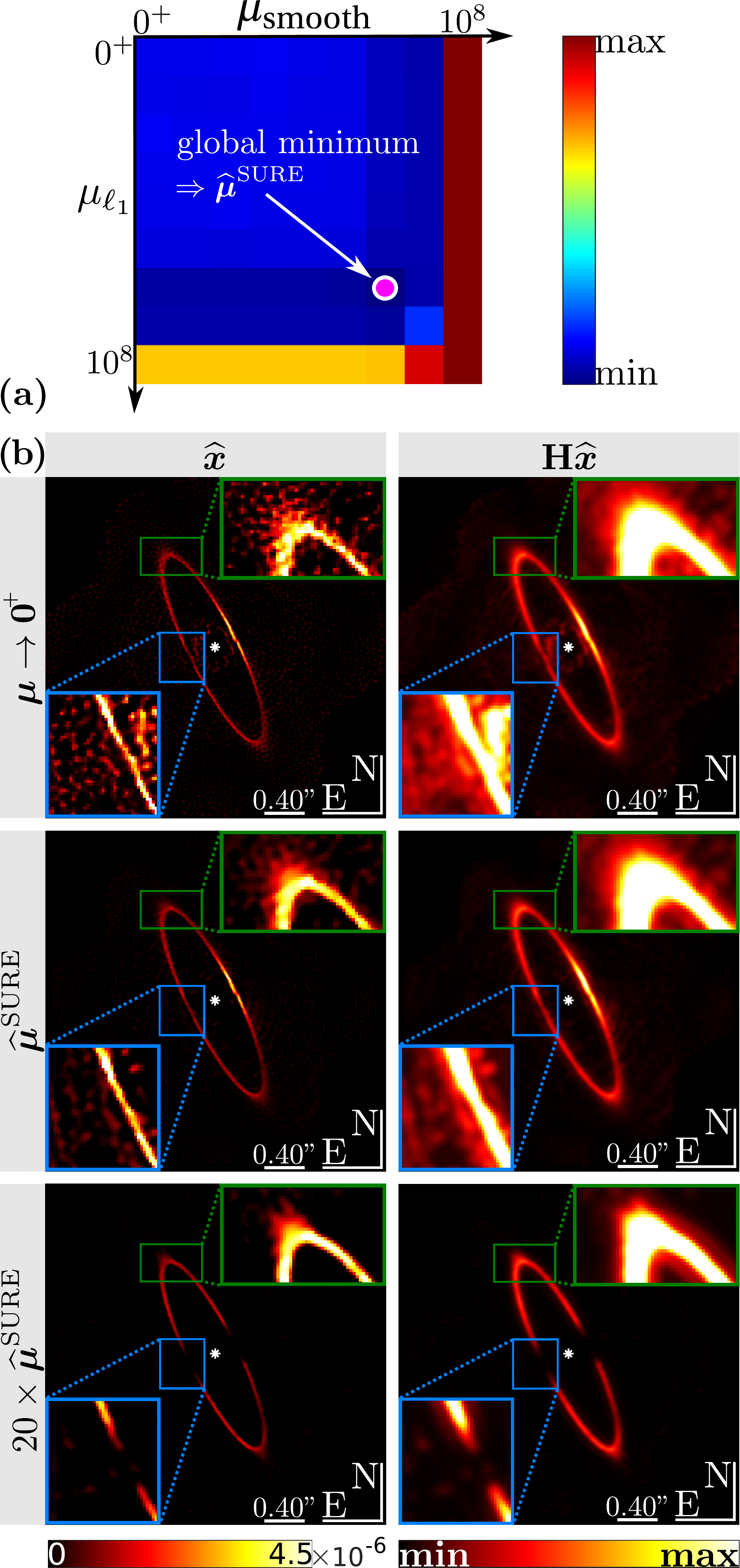}
			\caption{Effect of the setting of the
				regularization parameters on \REXPACO reconstructions. (a)
				SURE criterion derived from
				Eq.~(\ref{eq:patched_sure_criterion});
				(b) comparative reconstructions (first row: $\V \mu
				\to \M 0$, i.e. under-regularization;
				second row: $\estim{\V \mu}^{\text{SURE}} =
				\lbrace 10^{6}, 10^{6}\rbrace$, i.e. estimated optimal
				regularization; third row: $20\times \estim{\V
				\mu}^{\text{SURE}} =
				\lbrace 2\times10^{7}, 2\times10^{7}\rbrace$, i.e. over-regularization).
				Two regions of interest
				are also displayed with a saturated colormap near the north side
				extremity of the disk and near the star. Data set: HR 4796A, see
				Sect.~\ref{sec:datasets_description}
				for observing conditions.}
			\label{fig:regularization_fullfig}
		\end{figure}

		The comparison with \PACO is also interesting. While this latter algorithm
		demonstrates significantly better performance than the PCA to
		detect point-like sources \citep{flasseur2018exoplanet}, it performs worse
		than the PCA and cADI in the presence of extended features
		since it assumes that the pattern to detect is known and takes the form of a
		point-like source. Besides, parts of the disks are lost in the \PACO
		reconstructions due to the subtraction of the mean component of the ADI
		sequence\footnote{\PACO implements a dedicated strategy to cope
			with this issue for point-like sources: The candidate companions above a given
			threshold are characterized in terms of astrometry and photometry after their
			detection. This second step prevents a statistical bias in the estimated
			quantities.}. This effect is especially
		problematic for SAO 206462 since the nuisance
			component is hidden by the disk in almost all temporal frames due to the
		quasi-circular symmetry of the disk. The combination of the reconstruction
		framework of \REXPACO with its iterative strategy reducing bias
		of the statistics of the nuisance component leads to a
		physically more exploitable image of the flux distribution.
		The quality of the reconstruction is supported by the simulations
		performed in Sect.~\ref{sec:unbiased_estimation} (see
		Figs.~\ref{fig:hip80019_photometry_spiral_with_x_fullfig} and
		\ref{fig:hip80019_photometry_spiral_profiles_fullfig}) where a numerical model
		of a spiral disk with two arms of comparable spatial extent is injected to the
		data. These numerical tests have shown that \REXPACO is able to reconstruct a
		faithful image of the flux distribution of such a type of disk for levels of
		contrast $\gtrsim 5\times 10^{-6}$, that is to say about five to ten times fainter
		than the mean reconstructed level of flux of the SAO 206462 disk. The bright
		asymmetry in the southwest side of the disk could be interpreted as a
		reconstruction artifact possibly due to a local mis-modeling of the nuisance
		component, but we did not experience such an effect in our simulations for the
		typical level of flux of SAO 206462. Besides, the \REXPACO reconstruction
		exhibits detailed structures of the spiral arms with a better rejection of the
		nuisance component near the star than the ADI and RDI post-processings
		performed in \cite{maire2017testing} on the same sequence of observations.

		Finally, the case of PDS 70 is particularly interesting due to the complex
		structure of the disk and to the presence of two confirmed exoplanets at
		formation stage inside the disk. Concerning the disk structure, the outer ring
		reconstructed by \REXPACO is free from the self-subtraction artifacts that
		were visible on the cADI, PCA and \PACO reductions. The bright asymmetry of
		the disk in the west side of the outer disk and the possible spiral-shaped
		features pointed by \cite{muller2018orbital} in the north side of the disk are
		also detectable in the \REXPACO reconstruction. Concerning the point-like
		sources, the exoplanet PDS 70b exhibits a bright flux at $\lambda =
		2.11\,\micro\meter$ (K1 spectral band) as previously reported by
		\cite{muller2018orbital,mesa2019vlt}. Re-detecting the exoplanet PDS 70c is
		more difficult since it is embedded inside the disk material as reported by
		\cite{haffert2019two,mesa2019vlt}. We propose in Sect.~\ref{sec:results_joint}
		a processing strategy to disentangle the contribution of point-like sources
		from the spatially extended contribution of the circumstellar disk. Besides, a
		putative point-like feature (PLF) has been recently identified by
		\cite{mesa2019vlt} based on the analysis of several near-infrared
		angular plus spectral differential imaging (ASDI) sequences by various
		post-processing methods. This PLF can be (marginally) detected from the
		\REXPACO reconstruction while it is indistinguishable in the same ADI sequence
		processed by PCA (see Fig.~1 of \cite{mesa2019vlt}), as it is also the case
		for our PCA reduction given in Fig.~\ref{fig:reconstructions_fullfig}.

		Figure~\ref{fig:reconstruction_diagcov_fullfig} shows the importance of the
		modeling of spatial covariances in \REXPACO: Regularized inversions with a
		statistical model of the nuisance component that neglects all covariances (but
		still accounts for the nonstationary variances) are given in
		Fig.~\ref{fig:reconstruction_diagcov_fullfig}(a), to compare with \REXPACO
		reconstruction shown in Fig.~\ref{fig:reconstruction_diagcov_fullfig}(b). When
		spatial covariances are omitted, the reconstructions are worse. This is
		particularly visible on HR 4796A, in the background and close to the star, and
		on SAO 206462, close to the star.

		Figure~\ref{fig:regularization_fullfig} completes this study by showing that
		the method for automatically setting the regularization hyper-parameters
		discussed in Sect.~\ref{sec:automatic_setting_regularization_parameters} leads
		to satisfying reconstructions: The HR 4796A disk is smoother, with sharper
		edges, and the background is less noisy when automatically selected
		hyper-parameters $\widehat{\V \mu}^{\text{SURE}}$ are used
		compared to the case of under-regularization $\V \mu\to\V 0^+$. Conversely,
		over-regularization ($\V \mu = 20\times\widehat{\V
		\mu}^{\text{SURE}}$)
		leads to strong artifacts mimicking the self-subtraction
		effects, especially near the star, due to the large penalization of the
		$\ell_1$-norm of the reconstructed flux distribution. The extremities of the disk
		ring also exhibit staircase-artifacts. These effects are signs of an
		over-regularization by the edge-preserving term.

		\begin{figure*}[!t]
			\centering
			\includegraphics[width=\textwidth]{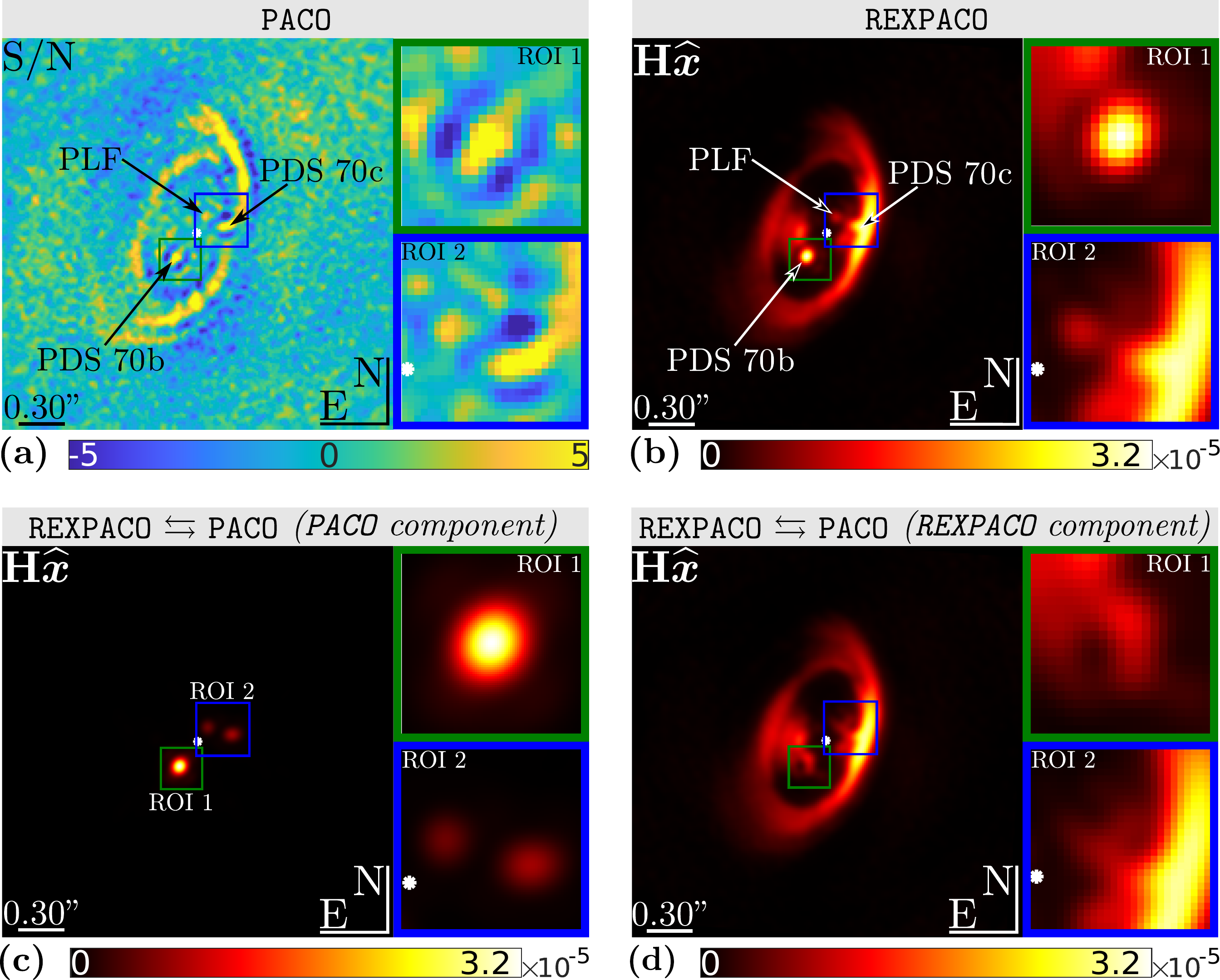}
			 \caption{Combining \REXPACO with
			\PACO on a data set of PDS 70 at $\lambda = 2.110\,\micro\meter$: (a) S/N map
			from \PACO; (b) flux distribution from \REXPACO; (c) contribution of the
			three
			selected point-like sources estimated with \PACO; (d) contribution of the
			disk reconstructed with \REXPACO.}
			\label{fig:pds70_rexpaco_paco_with_snr_fullfig}
		\end{figure*}
		
		\subsubsection*{Discussion of the results}
		\label{sec:discussion}

		Based on numerical injections of fake disks (with different
		orientations and typical morphologies), we have shown that \REXPACO is capable
		of reconstructing a faithful image of the flux distribution at contrasts up to
		$10^6$ for elliptical disks with a large eccentricity (about 0.95). For disks
		in face-on configuration, our simulations have shown that the contrast
		achieved by \REXPACO drops by a factor between two and ten due to the
		increased difficulty in disentangling the disk signal from the nuisance
		component in such a configuration. While these first quantitative results
		should be complemented and refined by more extensive simulations with diverse
		disks and observing conditions, they give first clues about the operating
		range of \REXPACO. In addition, \REXPACO better preserves both the morphology
		and the photometry of the disks than the standard cADI method used routinely
		to process ADI sequences in the presence of circumstellar disks.

		We have also applied \REXPACO on four ADI sequences recorded by the
		VLT/SPHERE-IRDIS instrument on stars hosting known circumstellar
		disks with various morphologies. While there is no ground truth for these
		objects, we have shown that \REXPACO copes with the classical ADI artifacts
		(e.g., nonphysical distortions and large residual stellar leakages).
		The images produced by \REXPACO allow to recover
		structures in the circumstellar material that were previously identified only
		with methods complementing ADI such as reference or polarization differential
		imaging. We expect that the high contrast capability of \REXPACO combined with
		its ability to reconstruct faithful images of the circumstellar material will
		be helpful in detecting new circumstellar disks and in refining the
		astrophysical interpretation of their flux distributions.

		\section{Unmixing point-like sources and extended sources}
		\label{sec:results_joint}

		\begin{figure*}[!t]
			\centering
			\includegraphics[width=\textwidth]{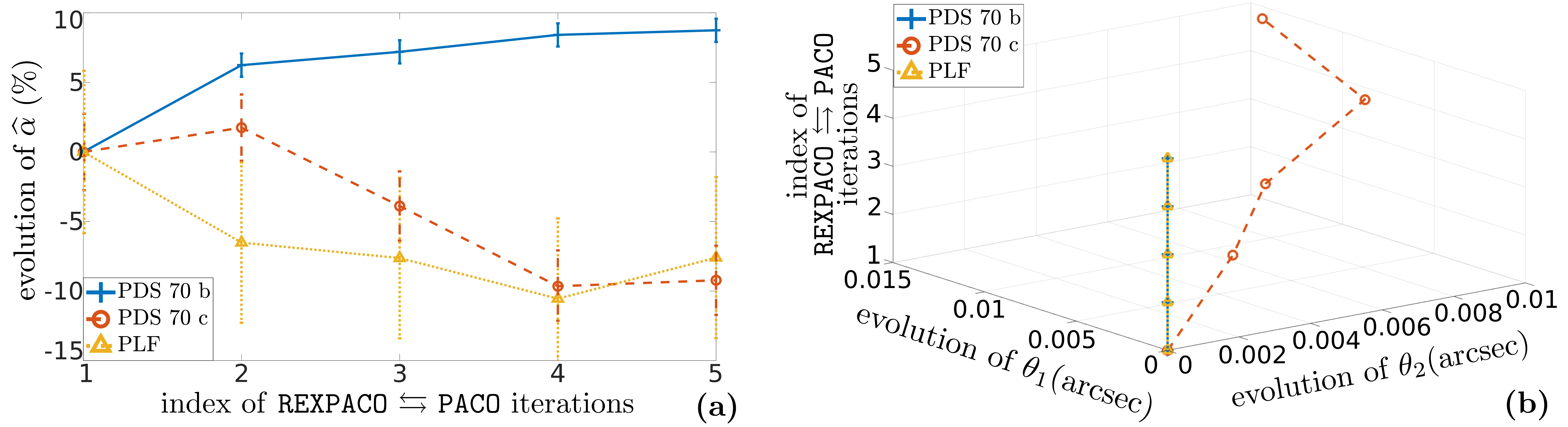}
			 \caption{Combining \REXPACO with
			\PACO on a data set of PDS 70 at $\lambda = 2.110\,\micro\meter$:  (a)
			evolution of the estimated photometry; (b) evolution of the estimated
			astrometry, for the three selected point-like sources.}
			\label{fig:pds70_characterization_fullfig}
		\end{figure*}

		For systems formed of both an extended disk and point-like
		sources such as exoplanets, the image reconstruction can be improved by
		combining \REXPACO algorithm with the exoplanet detection and characterization
		method \PACO. Both algorithms share a common statistical modeling of the
		background and can thus be combined in a principled way. In our
		experiments,
		this approach proved superior to our attempts to reconstruct with \REXPACO a
		composite object $\V x = \V x_\text{ext} + \V x_\text{pnt}$ where $\V
		x_\text{ext}$ accounts for extended (smooth) structures while $\V
		x_\text{pnt}$ accounts for point-like sources and with a regularization
		designed to favor these specific structures in each component:
		\begin{displaymath}
			\mathscr{R}(\V x, \V \mu) =
			\mu_{\text{smooth}}\,\mathscr{R}_{\text{smooth}}(\V x_\text{ext}, \epsilon) +
			\mu_{\ell_1}\,\mathscr{R}_{\ell_1}(\V x_\text{pnt})\,.
		\end{displaymath}
		The two main arguments in favor of combining \REXPACO and \PACO are the ability of
		\PACO to account for the sub-pixel location of point-like sources (while a
		sparse component $\V x_\text{pnt}$ is restricted to the discretization of
		the pixel grid) and direct control on the number of point-like sources (while
		$\V x_\text{pnt}$ either displayed many more nonzero entries than the
		actual number of point sources, or missed the weaker point-like
		sources).%

		In a nutshell, the proposed combination of the two algorithms works as
		follows: (i) \PACO and \REXPACO are launched independently on a given
		ADI sequence; (ii) based on the S/N map produced by \PACO and
		on the flux distribution produced by \REXPACO, the user
		identifies manually putative point-like sources; (iii)
		\REXPACO and \PACO are then alternatively applied. During step (iii), the
		astrometry and photometry of the selected point-like sources
		are refined by \PACO, based on the residual data obtained after subtraction of
		the disk contribution as currently reconstructed by \REXPACO. Next, the flux
		distribution of the disk can be refined by \REXPACO on updated
		residuals obtained after subtraction of the refined point-sources contribution
		estimated by \PACO. Step (iii) is repeated until convergence of the disk and
		point-like sources components. In
            practice, we stop the alternating procedure of step (iii) when the
            relative
            evolution of both the disk and the point-like sources components
            is smaller
            than $10^{-8}$ between two consecutive iterations.
		This strategy can be particularly valuable for the processing of ADI
		sequences of stars that host point-like sources near a
		circumstellar disk. In this paper, we illustrate our combined
		approach on the PDS 70 system hosting two confirmed exoplanets (PDS 70b and
		PDS 70c) inside a protoplanetary disk
		\citep{keppler2018discovery,haffert2019two}, see
			Sect.~\ref{sec:results_reconstruction_disk} for a discussion about the
			\REXPACO reconstruction. Based on the \PACO S/N map and \REXPACO flux
		distribution provided by step (i) of the combined method, we
		manually selected the rough location of the exoplanets PDS 70b and PDS 70c, see
		Fig.~\ref{fig:pds70_rexpaco_paco_with_snr_fullfig}(a)-(b). We
		also considered the point-like feature (PLF) first identified by
		\cite{mesa2019vlt}\footnote{Based on estimated photometry, the study of
			\cite{mesa2019vlt} concludes that this PLF is most likely a part of the disk.
			In our present work, the goal is not to discuss the status of the PLF but
			rather to illustrate the principle of the proposed algorithm.}, see
		Fig.~\ref{fig:pds70_rexpaco_paco_with_snr_fullfig}(a)-(b).
		Figure~\ref{fig:pds70_rexpaco_paco_with_snr_fullfig}(c)-(d) shows the
		contribution of the three selected point-like features (PDS 70b, PDS 70c and
		the PLF) estimated by \PACO and the disk contribution reconstructed by
		\REXPACO after convergence of the proposed alternating scheme.
		Figure~\ref{fig:pds70_rexpaco_paco_with_snr_fullfig}(c)-(d) shows the good
		separation of point-like sources and of the extended disk structures surrounding
		them, thus illustrating the efficiency of this approach to disentangle the sparse
		contribution of the point-like sources from the spatially
		extended contribution of the disk. It is worth noticing that near the location
		of PDS 70b, the spatially extended component retrieved by the
		\REXPACO/\PACO combination exhibits an extended faint circular structure that
		could be a hint
		of a circumplanetary disk as proposed by \cite{Christiaens2019} and \cite{Isella2019}.
		It is not yet possible to conclude on the reality of this
		structure recovered by \REXPACO/\PACO, which could also
		be the result of reconstruction artifacts (e.g., due to a partial
		unmixing of the sparse and spatially extended components). This open
		question would require further analysis, for instance based on numerical simulations
		by means of injection of fake exoplanets and of a physics-based model of the
		PDS 70 disk to study the influence of the exoplanet-disk interactions on the
		reconstruction results. Such a study is left for future
		work. Concerning PDS 70c, the
		extended feature detected in the disk component near this exoplanet is in good
		agreement with the submillimeter continuum emission detection
		\citep{Isella2019} which confirms the hypothesis that this exoplanet is
		still in its accretion phase.

		Figure~\ref{fig:pds70_characterization_fullfig} shows how the photometry and
		astrometry of the three point sources evolve throughout the alternations
		between \REXPACO and \PACO. Accounting for the presence of the disk modifies
		the estimated photometry of the three selected
		point-like sources by about 10\%. The astrometry of PDS 70b and
		of the PLF estimated with \PACO are not significantly modified by the proposed
		alternating scheme: These two point-like sources are not located inside a
		bright arc of the disk so that the disk material introduces only a small bias
		on \PACO estimations.  On the other hand, the estimated astrometry of the
		exoplanet PDS 70c located inside a very bright arm of the disk is modified by
		about 0.010 arcsec (i.e., about 1 pixel in the two spatial directions, \PACO
		astrometry is evaluated on a subpixel grid with four nodes per pixel). The
		astrophysical interpretation of these results would require a dedicated study
		on multiple epochs. If the astro-photometric estimations we derived are
		confirmed, they could be significant corrective factors
		of the spectral energy distribution and of the orbit of the exoplanet
		PDS 70c.

		\section{Conclusion} \label{sec:conclusion}

		In this paper, we have introduced a new post-processing
		algorithm, named \REXPACO, dedicated to the reconstruction of flux
		distributions from ADI data sets in the presence
		of extended features, like circumstellar disks. To the best of
		our knowledge, \REXPACO is one of the first methods specifically designed for
		this task. It encompasses the statistical modeling of the fluctuations
		of the nuisance component (i.e., stellar leakages and
		additional sources of noise) at the scale of small patches
		of a few tens of pixels. The resulting statistics of
		the nuisance term are accounted for in a regularized
		reconstruction framework following an inverse problem
		approach based on a forward image formation model of the
		off-axis sources in the ADI sequences. We have shown that suitable spatial
		regularizations ($\ell_1$ norm and $\ell_2 - \ell_1$
		edge-preserving) reduce residual reconstruction artifacts.
		Regularization parameters are estimated by minimization of
		SURE. Our numerical experiments have shown that
		this strategy leads to a quasi-optimal setting of these parameters. The
		images of the circumstellar environment produced by our
		algorithm are deblurred from the instrumental PSF
		which is accounted for in the forward model of the image
		formation. These different ingredients make our approach very
		versatile and capable of recovering both sharp edges and smooth transitions of
		the circumstellar material. The \REXPACO algorithm is fully
		unsupervised: The patch size, the statistics of the nuisance component and the
		weights given to the regularizations are estimated in a data-driven fashion.
		Besides, we have illustrated the benefits of the combination of \REXPACO with
		\PACO, two algorithms dedicated to complementary tasks, to analyze ADI
		sequences in the presence of both circumstellar disks and
		candidate point-like sources. The combination of
		the two algorithms is able to disentangle the spatially extended contribution of
		circumstellar disks from the spatially localized contribution of point-like
		sources.

		While the statistical model of \REXPACO is able to capture most of the
		fluctuations of the nuisance component, slight stellar residuals are still
		present in the reconstructions. The fidelity of the underlying statistical
		model to the empirical distribution of the nuisance component is a key point
		to extract accurately the signal from off-axis sources. Beyond
		accurate modeling of the nuisance term, the angular motion through the ADI
		sequence is also a key point to unmix the light
		coming from off-axis objects such as a circumstellar disk
		and point-like sources from the stellar light that leaks from
		the coronagraph. For instance, objects with a circular symmetry cannot be
		correctly recovered since the temporal diversity brought by ADI
		is not sufficient in this case. Only additional information such as the
		spectral measurements obtained with angular plus spectral
		differential imaging (ASDI) in addition to the temporal diversity of ADI
		would make the reconstruction possible in those cases. The
		refinement and adaptation of \REXPACO to the two above-mentioned points
		require specific refinements that will be developed in a future paper.

		\section*{Acknowledgments}

		We thank A. Boccaletti (LESIA, Paris, France) who provided the transmission of
		the SPHERE coronagraphs. We also thank the anonymous referee for
		her/his careful reading of the manuscript as well as her/his insightful
		comments and suggestions.

		This work has made use of the SPHERE Data Center, jointly operated by
		OSUG/IPAG (Grenoble, France), PYTHEAS/LAM/CESAM (Marseille, France),
		OCA/Lagrange (Nice, France), Observatoire de Paris/LESIA (Paris, France), and
		Observatoire de Lyon/CRAL (Lyon, France).

		This work has been supported by the Région Auvergne-Rhône-Alpes
		under the project DIAGHOLO, by the French National Programs
		(PNP and PNPS), and by the Action Spécifique Haute Résolution Angulaire
		(ASHRA) of CNRS/INSU co-funded by CNES.

		\bibliographystyle{aa} 
		\bibliography{REXPACO}

		\appendix

		\section{Constrained minimization of the cost function}
		\label{app:VMLMB}

		The restored object $\V x$ is the minimum of the objective function
		$\mathscr{C}(\V r, \V x, \M A, \M \Omega, \V \mu)$ defined in
		Eq.~\eqref{eq:global_argmin} on the nonnegative orthant (i.e., $\V x$ is
		everywhere nonnegative). We numerically solve this constrained optimization
		problem with the VMLM-B algorithm \citep{thiebaut2002optimization} as
		implemented in
		\textsc{OptimPackLegacy}\footnote{\url{https://github.com/emmt/OptimPackLegacy}}.
		VMLM-B is a limited memory quasi-Newton method implementing optional bound
		constraints on the sought parameters. This method has proven to be very effective
		for solving large-scale (millions or even billions of variables) nonlinear
		problems with or without bound constraints that are typical of inverse problems
		for image restoration.  The requirements for running VMLM-B are to provide the
		bounds (if any) and a numerical function to compute the objective function and
		its gradient for a given set of parameters $\V x$.

		The gradient of the objective function $\mathscr{C}(\V r, \V x, \M A, \M
		\Omega, \V \mu)$ with respect to $\V x$ is given by:
		\begin{multline} {\boldsymbol{\nabla}}_{\V x}\,\mathscr{C}(\V r, \V x, \M A,
			\M \Omega, \V \mu) = \\ \underbrace{\sum\limits_{n \in \DisjointPatches}
			\sum\limits_{t=1}^T \M A_t\T \, \M P_n\T \left[ \ShrunkCov_n \left( \M P_n
			\left( \M A_t\,\V x - \V r_t + \estim{\V m} \right) \right) \right]}_{=
			\boldsymbol{\nabla}_{\V x}\, \mathscr{D}} \\ + \mu_{\text{smooth}}\,
			\underbrace{\frac{\M \Delta_n\T \, \M \Delta_n \, \V x}{\sqrt{||\M \Delta_n \,
			\V x||_2^2 + \epsilon^2}}}_{=\boldsymbol{\nabla}_{\V
			x}\,\mathscr{R}_{\text{smooth}}} + \, \mu_{\ell_1}\hspace{-2mm}\underbrace{\M
			1_N}_{\boldsymbol{\nabla}_{\V x}\, \mathscr{R}_{\ell_1}}\,.
			\label{eq:gradient_cost_function}
		\end{multline}%
		The computational burden for the gradient is similar to that of the objective
		function.

		\section{Importance of removing off-axis sources when learning the
		statistical model of the nuisance component}
		\label{app:impcov}

		\begin{figure*}[t!] \centering%
			\includegraphics[width=0.9\textwidth]{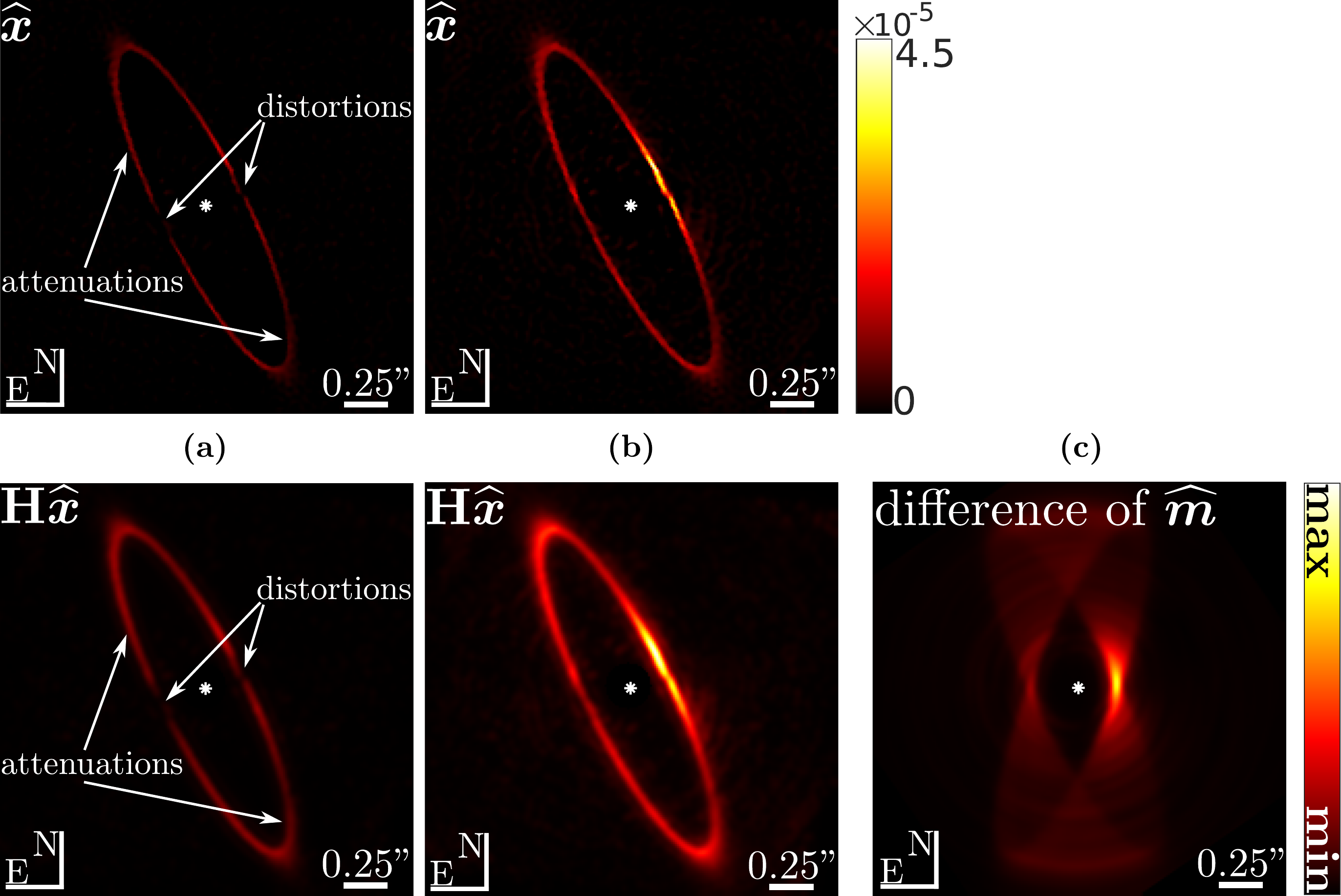}%
			\caption{%
				Illustration of the influence of the statistics $\estim{\M \Omega} =
				\lbrace
				\estim{\V m}, {\lbrace \ShrunkCov_n \rbrace}_{n \in \DisjointPatches}
				\rbrace$ of the nuisance component on the
					\REXPACO reconstructions. (a) $\estim{\M \Omega}$ is computed with
				Eqs.~\eqref{eq:estimator_sample_covs} ignoring the
				contribution of the object; (b) $\estim{\M \Omega}$ is computed with
				Algorithm~\ref{alg:update_stat}; (c) difference of the mean component
				$\estim{\V m}$ estimated with
				Eq.~(\ref{eq:estimator_sample_covs}a) and Algorithm
				\ref{alg:update_stat}. For (a) and (b), both the deconvolved flux
				distribution $\widehat{\V x}$  and its convolved version $\M
				H\,\widehat{\V
					x}$ are given. The difference of estimated means $\widehat{\V m}$
				given in
				(c) is represented with the same colorbar than the convolved objects $\M
				H\,\widehat{\V x}$ given in (a) and (b) to ease comparisons. Data set: HR
				4796A, see Sect.~\ref{sec:datasets_description} for observing
				conditions.}%
			\label{fig:mean_fullfig}
		\end{figure*}

		Figure~\ref{fig:mean_fullfig}(a) illustrates on the example of HR 4796A that
		estimating the statistics $\M \Omega$ of the nuisance component directly from
		the data with Eqs.~\eqref{eq:estimator_sample_covs} leads to self-subtraction
		artifacts: A portion of the light of off-axis sources gets removed from the
		measurements because it is erroneously attributed to the nuisance component
		(the mean $\estim{\V m}^{[1]}$ defined in
		Eq.~(\ref{eq:estimator_sample_covs}a)).

		Figure~\ref{fig:mean_fullfig}(b) shows that refining the statistics of the
		nuisance component to compensate for the contribution of the off-axis sources
		improves the reconstruction: Parts of the disk (like its bright west side) that
		were previously missed are covered.
		Figure~\ref{fig:mean_fullfig}(c) displays
		the difference between the mean component $\estim{\V m}$ estimated with
		Eq.~(\ref{eq:estimator_sample_covs}a) and with Algorithm~\ref{alg:update_stat}.
		This difference is largest near the star where a significant part of disk
		contribution is lost in the average of the nuisance component $\estim{\V m}$
		when using Eq.~(\ref{eq:estimator_sample_covs}a). This explains the bias and
		artifacts observed in Fig.~\ref{fig:mean_fullfig}(a).

		\section{Hierarchical estimation of the statistics of the nuisance component and
		of the flux distribution}
		\label{app:hierarchical}

		We first recall the expression of the shrinkage estimator $\widehat{\M \Omega}
		= \lbrace \widehat{\V m}, {\lbrace  \ShrunkCov_n \rbrace}_{n \in \mathbb{P}}
		\rbrace$ of the nuisance component given in Eqs.~(\ref{eq:estimator_m}) to
		(\ref{eq:shrinkage_convex_combination_W}):
		\begin{equation}
			\begin{cases}
				\widehat{\V m}(\V x) = \frac{1}{T}\sum\limits_{t=1}^T \left(\V r_t - \M
				A_t \, \V x\right) \vspace{0.5mm} \\
				\ShrunkCov_n(\V x) = \widetilde{\M W}_n \odot \SampleCov_n(\V x)
				\,,\forall n\in\mathbb{P}
			\end{cases}
			\label{eq:shrunk_statistics}
		\end{equation}
		with the empirical covariance matrix defined by:
		\begin{equation}
			\SampleCov_n(\V x) = \frac{1}{T} \sum_{t=1}^{T} \estim{\V u}_{n,t}(\V x)\,
			\estim{\V u}_{n,t}(\V x)\T\,,
		\end{equation}
		and the residuals
		\begin{equation}
			\estim{\V u}_{n,t}(\V x) = \M P_{n}\, \Paren*{\V r_t - \estim{\V m}(\V x) -
			\M A_t \,{\V x}}\,.
		\end{equation}
		The shrinkage matrix $\widetilde{\M W}_n$ is defined by
		Eq.~(\ref{eq:W}) and is considered fixed (i.e., independent from $\V x$).
		In the following we show that these estimators are minimizers of the modified cost function
		$\mathscr{D}_{\text{joint}}$ defined in
		Eq.~(\ref{eq:co-log-likelihood_shrinkage}):
		\begin{equation}
			\left(\widehat{\V m}(\V x),\left\{\ShrunkCov_n(\V
			x)\right\}_{n\in\mathbb P}\right) = \argmin_{\V m, \{\M C_n\}}
			\mathscr{D}_{\text{joint}}(\V r, \V x, \V m, \{\M C_n\}_{n\in\mathbb{P}})
		\end{equation}
		with
		\begin{multline}
			\mathscr{D}_{\text{joint}}(\V r, \V x, \V m, \{\M
			C_n\}_{n\in\mathbb{P}}) = \frac{T}{2} \sum\limits_{n \in \DisjointPatches}
			\log \det{\M C_n} \\
			+\frac{1}{2} \sum\limits_{n \in \DisjointPatches}  \Trace
			\Biggl( \M C_n^{-1} \Biggl( \widetilde{\M W}_n \odot \sum_{t=1}^{T} \V
			u_{n,t}\,\V u_{n,t}\T \Biggr)\Biggr)\,,
			\label{eq:co-log-likelihood_shrinkage2}
		\end{multline}
	and
	\begin{equation}
		\V u_{n,t} = \M P_{n}\, \Paren*{\V r_t - \V m -
		\M A_t \,{\V x}}\,.
	\end{equation}
	\begin{proof}
		For a fixed $\V x$, the differentiation of
		$\mathscr{D}_{\text{joint}}$ with respect to $\M \Omega_n = \lbrace \V m, \M
		C_n \rbrace$ leads to:
		\begin{multline}
			\text{d}\mathscr{D}_{\text{joint}}\Big|_{\M \Omega_n} =
			\frac{T}{2}\Trace\left( \M C_n^{-1} \, \text{d}\M C_n \right) + \\
			\frac{1}{2}\Trace\Bigg[-2\M C_n^{-1}\Bigg( \widetilde{\M W}_n \odot
			\sum\limits_{t=1}^T \V u_{n,t}\,(\text{d}\V m)\T \Bigg) \\ - \M
			C_n^{-1}\,\text{d}\M C_n\,\M C_n^{-1}\Bigg( \widetilde{\M W}_n \odot
			\sum\limits_{t=1}^T \V u_{n,t} \, \V u_{n,t}\T \Bigg) \Bigg]\,.
		\end{multline}
		The necessary first-order optimality condition:
		\begin{displaymath}
			\text{d}\mathscr{D}_{\text{joint}}\Big|_{\M \Omega_n}=0
		\end{displaymath}
		leads to:
		\begin{equation}
			\begin{cases}
				\M C_n^{-1}\Bigg( \widetilde{\M W}_n
				\odot \sum\limits_{t=1}^T \V u_{n,t} \Bigg) = \M 0 \\
				T \, \M I_{K} - \M C_n^{-1}\Bigg( \widetilde{\M W}_n \odot
				\sum\limits_{t=1}^T \V u_{n,t} \, \V u_{n,t}\T \Bigg) = \M 0\,,
			\end{cases}
			\label{eq:optimality_cond_system}
		\end{equation}
		where $\M I_{K} \in \mathbb{R}^{K\times K}$ stands for the identity matrix.
		Since $\M C_n$ is necessary non-singular, the system of
		equations~(\ref{eq:optimality_cond_system}) gives:
		\begin{equation}
			\begin{cases}
				\V m = \frac{1}{T}\sum\limits_{t=1}^T (\V r_t - \M A_t \, \V x) \\
				\M C_n = \widetilde{\M W}_n \odot \frac{1}{T} \sum\limits_{t=1}^T \V
				u_{n,t}
				\,
				\V u_{n,t}\T\,,
			\end{cases}
		\end{equation}
		which corresponds to the general form of the estimators given in
		Eqs.~(\ref{eq:shrunk_statistics}).
	\end{proof}

\end{document}